\documentclass[%
reprint,
amsmath,amssymb,
aps,superscriptaddress
]{revtex4-2}
\allowdisplaybreaks
\usepackage{graphicx}
\usepackage{dcolumn}
\usepackage{bm}
\usepackage{multirow}
\usepackage{amssymb}
\usepackage[
bookmarks=true,
colorlinks,
linkcolor=blue,
urlcolor=blue,
citecolor=blue,
plainpages=false,
pdfpagelabels,
final,
breaklinks=true
]{hyperref}

\begin{document}
	\bibliographystyle{unsrt}
	\preprint{APS/123-QED}
	
	\title{Strong nonlocal sets of UPB}
	
	\author{Bichen Che}
	\affiliation{Information Security Center, State Key Laboratory of Networking and Switching Technology, Beijing University of Posts and Telecommunications, Beijing 100876, People's Republic of China}%
	\affiliation{Information Security Center, Beijing University of Posts and Telecommunications, Beijing 100876, People's Republic of China}%
	\author{Zhao Dou}
	\email{dou@bupt.edu.cn}
	\affiliation{Information Security Center, State Key Laboratory of Networking and Switching Technology, Beijing University of Posts and Telecommunications, Beijing 100876, People's Republic of China}%
	\author{Min Lei}
	\affiliation{Information Security Center, Beijing University of Posts and Telecommunications, Beijing 100876, People's Republic of China}
	\author{Yixian Yang}
	\affiliation{Information Security Center, State Key Laboratory of Networking and Switching Technology, Beijing University of Posts and Telecommunications, Beijing 100876, People's Republic of China}%

	
	
	

\begin{abstract}
The unextendible product bases (UPBs) are interesting members from the family of orthogonal product states. In this paper, we investigate the construction of 3-qubit UPB with strong nonlocality of different sizes. First, a UPB set in ${{C}^{3}}\otimes {{C}^{3}}\otimes {{C}^{3}}$ of size 12 is presented based on the Shifts UPB, the structure of which is described by mapping the system to a $3\times 3\times 3$ Rubik's Cube. After observing the orthogonal graph of each qubit, we provide a general method of constructing UPB in ${{C}^{d}}\otimes {{C}^{d}}\otimes {{C}^{d}}$ of size ${{\left( d-1 \right)}^{3}}+3\left( d-2 \right)+1$. Second, for the more general case where the dimensions of qubits are different, we extend the tile structure to 3-qubit system and propose a Tri-tile structure for 3-qubit UPB. Then, by means of this structure, a ${{C}^{4}}\otimes {{C}^{4}}\otimes {{C}^{5}}$ system of size 30 is obtained based on a ${{C}^{3}}\otimes {{C}^{3}}\otimes {{C}^{4}}$ system. Similarly, we generalize this approach to ${{C}^{{{d}_{1}}}}\otimes {{C}^{{{d}_{2}}}}\otimes {{C}^{{{d}_{3}}}}$ system which has a similar composition to ${{C}^{d}}\otimes {{C}^{d}}\otimes {{C}^{d}}$. Our research provides a positive answer to the open questions raised in [Halder, et al., PRL, 122, 040403 (2019)], indicating that there do exist multi-qubit UPBs that can exhibit strong quantum nonlocality without entanglement.      
\end{abstract}

\maketitle


\section{\label{sec:level1}Introduction}

With its potential application value, quantum information has brought new changes to the information industry\cite{1,2,3,4}. Entanglement between quantum states is a kind of quantum resource, which can be used to achieve tasks that cannot be accomplished by classical resources\cite{5,6,7,8}, such as quantum teleportation\cite{9,10}, quantum algorithm\cite{11,12,13}, and quantum dense coding\cite{14}. For a long time, it has been believed that the nonlocality of quantum entanglement leads to these properties of entangled states. However, in 1999, Bennett et al.\cite{15} proposed a set of orthogonal product states with nonlocality, which aroused a wide discussion on the relationship between entanglement and nonlocality.

Unextendible product bases (or UPB) is a class of incomplete orthogonal product states, whose complementary space does not contain any product state \cite{16,17,18}. It cannot be distinguished perfectly by local operations and classical communication (LOCC). Ref. \cite{19,20,21} has shown that it can be used in the production of bound entangled (BE) states and some special bipartite entangled states that remain positive under partial transpose (PPT). Nevertheless, most of the current efforts are devoted to the construction of 2-qubit UPBs, while little progress has been made on multi-qubit UPBs \cite{22,23,24,25,26}. Chen et al.\cite{22} investigated the minimum size of UPB with local dimension equals 2, and analyzed the proposed sets using orthogonal graphs. Bej et al.\cite{23} proposed that a set of high-dimensional reducible unextendible product states can be obtained by adding several orthogonal product states to a set of low-dimensional unextendible product states in bipartite systems. Recently, a method to construct UPBs of different large sizes in ${{C}^{m}}\otimes {{C}^{n}}$ was put forward by Shi et al. \cite{24}, which uses \textit{U}-tile structure. Multi-qubit UPBs are not only valuable in quantum circuit construction and cryptography experiments, but also often used to construct tight Bell inequalities without quantum violations \cite{27}. Therefore, a general construction method for constructing a multi-qubit UPB is needed, which is the first idea (motivation) of this paper.

As one of the research hotspots in quantum information theory, the quantum state discrimination problem is the basis of other quantum problems \cite{28,29,30}. The distinguishable quantum states can be applied to some quantum information processing tasks, such as distributed quantum computing, while the indistinguishable quantum states are very common in the design of quantum cryptography protocols. Bennett et al.\cite{31} proposed that any set of orthogonal product states in $2\otimes N$ is distinguishable by LOCC. Zhang et al.\cite{32} gave a general method to construct indistinguishable multipartite orthogonal product states in ${{C}^{{{d}_{1}}}}\otimes {{C}^{{{d}_{2}}}}\otimes \cdots \otimes {{C}^{{{d}_{n}}}},\left( {{C}^{{{d}_{1,2,\cdots ,n}}}}\ge 3,n\ge 4 \right)$. In addition to the above work, Halder et al. \cite{33} recently found that in some nonlocal tripartite systems, when two parties are measured together, there is a possibility to distinguish some certain states of the set. Hence, the concept of strong nonlocality has been proposed and widely discussed. Inspired by Halder’s work, Zhang et al. \cite{34} presented two sets of quantum states with strong nonlocality in ${{C}^{3}}\otimes {{C}^{3}}\otimes {{C}^{3}}$ and ${{C}^{3}}\otimes {{C}^{3}}\otimes {{C}^{3}}\otimes {{C}^{3}}$. Shi et al. \cite{35} provided the process of constructing a set consisting of entangled states in ${{C}^{d}}\otimes {{C}^{d}}\otimes {{C}^{d}}$ system based on the Rubik's cube. However, there is no relevant research on the construction of multi-qubit UPB with strong nonlocality, which is the second idea (motivation) of this paper.

In addition, graph theory is an effective way to show the abstract structure of state sets intuitively \cite{36,37}. It is also widely used in the construction of quantum state sets \cite{38,39,40}, especially UPBs\cite{41,42,43}. Johnston et al. \cite{37} analyzed the structure of UPB in ${{\left( {{C}^{2}} \right)}^{\otimes p}}$ and proposed the minimum size of this quantum system by using the orthogonal graph. Bennett et al. \cite{31} first provided two classical construction methods for UPB construction, namely Pyramid structure and tile structure. Hence, our third motivation is to use graph theory to better analyze and display the internal structure and relations of the strongly nonlocal states.

Therefore, on account of the aforesaid three motivations, the main focus in this paper is to construct a general set of 3-qubit UPB with strong nonlocality. First, based on the Shifts UPB, a UPB set in ${{C}^{3}}\otimes {{C}^{3}}\otimes {{C}^{3}}$ is obtained. Geometrically, by observing the orthogonal graph of each qubit and the corresponding $3\times 3\times 3$ Rubik's cube, the structure of the states set and the relationship between each qubit are analyzed in detail in Fig. 3 and Fig. 4. Then, following this construction method, we construct a general 3-qubits UPB with strong nonlocality of size ${{\left( d-1 \right)}^{3}}+3\left( d-2 \right)+1$ by dividing a $d\times d\times d$ Rubik's cube, and also give the general expression of the states set. Second, after reviewing the connection between UPBs and tile structure, we extend the tile structure from 2-qubit to 3-qubit systems and introduce it in Definition 5. Moreover, by applying Proposition 1, we generalize the construction of ${{C}^{4}}\times {{C}^{4}}\times {{C}^{5}}$ and show a universal approach to construct a strongly nonlocal UPB set with high-dimensional, ${{C}^{{{d}_{1}}}}\otimes {{C}^{{{d}_{2}}}}\otimes {{C}^{{{d}_{3}}}}$, based on a UPB set with low-dimensional,  ${{C}^{\left( {{d}_{1}}-1 \right)}}\otimes {{C}^{\left( {{d}_{2}}-1 \right)}}\otimes {{C}^{\left( {{d}_{3}}-1 \right)}}$. Our research on the general construction and discrimination of UPB can be applied to many practical quantum protocols, because the security of protocols is guaranteed fundamentally.

The rest of this paper is organized as follows. In Sec. II, we briefly introduce the notations and several preliminary concepts of UPBs. Sec. III and Sec. IV consist of the main contributions of the present work. Based on graph theory, the general construction method of three-qubit UPB with the same dimension and different dimensions are proposed respectively. Finally, we summarize the results and discuss some open problems in Sec. V.

\section{\label{sec:level2}Notations and preliminaries}
In this section, we briefly introduce the preliminary knowledge and some notations. A multi-qubit pure state $\left| v \right\rangle \in {{C}^{{{d}_{1}}}}\otimes \cdots \otimes {{C}^{{{d}_{p}}}}$, is considered to be separable if and only if it can be written in the form
\[\left| v \right\rangle =\left| {{v}_{1}} \right\rangle \otimes \cdots \otimes \left| {{v}_{p}} \right\rangle . \]

The standard bases of ${{C}^{{{d}_{{}}}}}$ is $\left\{ \left| 0 \right\rangle ,\ \left| 1 \right\rangle ,\ \cdots ,\ \left| d-1 \right\rangle  \right\}$. The symbol of the tensor product is sometimes omitted in order to discuss multiple qubit states more clearly. Note that throughout this paper, the states and operators are not normalized for simplicity, and only pure states and positive operator-valued measure (POVM) measurements are considered.

\textbf{Definition 1: } Consider a \textit{p} partite quantum system $  \mathcal {H} =\otimes _{i=1}^{p}{{ \mathcal {H} }_{i}}$. An orthogonal product bases (PB) is a set \textit{S} of pure orthogonal product states spanning a subspace ${{\mathcal {H} }_{S}}$ of $\mathcal {H}$. An uncompletable product bases (UCPB) is a PB whose complementary subspace $ \mathcal {H} _{S}^{\bot }$ contains fewer mutually orthogonal product states than its dimension. An unextendible product bases (UPB) is an uncompletable product bases that does not contain any product state in complementary subspace $ \mathcal {H} _{S}^{\bot }$.

UPB is nonlocal and cannot be perfectly distinguished by LOCC. But when discussing the nonlocality of a multi-qubit system, it is found that there is a certain probability that states can be distinguished when several qubits are joined. Based on this phenomenon, the definition of strong nonlocality is given.

\textbf{Definition 2: } In a multiparty system ${{C}^{{{d}_{1}}}}\otimes {{C}^{{{d}_{2}}}}\otimes \cdots \otimes {{C}^{{{d}_{n}}}},\ \left( {{d}_{1,2,\ \cdots ,\ n}}\ge 3,\ n\ge 4 \right)$, if a set of orthogonal product states is arbitrarily divided into \textit{i} parts, and the entire system is still locally irreducible in every new \textit{i} parts, the system is called \textit{i}-divisible, $i=2,3,\ \cdots ,\ (n-1)$.

\textbf{Definition 3: } In ${{C}^{{{d}_{1}}}}\otimes {{C}^{{{d}_{2}}}}\otimes \cdots \otimes {{C}^{{{d}_{n}}}},\ \left( {{d}_{1,2,\ \cdots ,\ n}}\ge 3,\ n\ge 4 \right)$, if a set of orthogonal product states is $(n-1)$-divisible, $(n-2)$-divisible...and 2- divisible simultaneously, it is said that this system is strongly nonlocal.

For the measurement of nonlocal state sets, since multiple rounds of measurement are required for multiple participants, it is necessary to carry out the nontrivial orthogonality-preserving measurement to obtain useful information without affecting the characteristics of the state set, which is defined in Definition 4.

\textbf{Definition 4: } If a set of mutually orthogonal quantum states remains mutually orthogonal after measurement, the measurement used to distinguish the quantum states is defined as orthonormal preserving measurement (OPM). Furthermore, such a measurement is called nontrivial if all the measurement matrices constituting the OPM are not proportional to the identity operator, otherwise, it is trivial.

Tile structure is one of the classical structures used to construct UPB. The ${{C}^{m}}\otimes {{C}^{n}}$ system can correspond to an $m\times n$ rectangle $\Gamma $, which is paved by disjoint tiles $\left\{ {{t}_{i}} \right\}$, denoted by $\Gamma =\bigcup\nolimits_{i}{{{t}_{i}}}$. A tile ${{t}_{i}}$ should be a rectangle that can be separated. In particular, we show how to construct a 2-qubit UPB of size 5 in Fig. 1.


\begin{figure}[b]
	\includegraphics[width=0.2\textwidth]{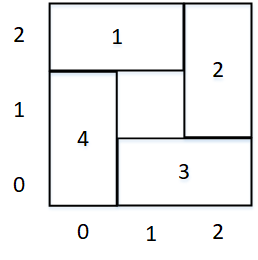}
	\caption{\label{fig:epsart} Tile structure.}
\end{figure}

Example 1: The following five states form a UPB in ${{C}^{3}}\otimes {{C}^{3}}$ system denoted the tile structure.

From Fig.1, we obtain a set of complete orthogonal product bases as Eq. (1), which denoted as  :

\begin{equation}
	\begin{aligned}
		& \left| \psi _{0}^{\left( 1 \right)} \right\rangle =\frac{1}{\sqrt{2}}\left| 0 \right\rangle \left( \left| 0 \right\rangle +\left| 1 \right\rangle  \right),\quad \left| \psi _{0}^{\left( 2 \right)} \right\rangle =\frac{1}{\sqrt{2}}\left| 0 \right\rangle \left( \left| 0 \right\rangle -\left| 1 \right\rangle  \right) \\ 
		& \left| \psi _{1}^{\left( 1 \right)} \right\rangle =\frac{1}{\sqrt{2}}\left( \left| 0 \right\rangle +\left| 1 \right\rangle  \right)\left| 2 \right\rangle ,\quad \left| \psi _{1}^{\left( 2 \right)} \right\rangle =\frac{1}{\sqrt{2}}\left( \left| 0 \right\rangle -\left| 1 \right\rangle  \right)\left| 2 \right\rangle , \\ 
		& \left| \psi _{2}^{\left( 1 \right)} \right\rangle =\frac{1}{\sqrt{2}}\left| 2 \right\rangle \left( \left| 1 \right\rangle +\left| 2 \right\rangle  \right),\quad \left| \psi _{2}^{\left( 2 \right)} \right\rangle =\frac{1}{\sqrt{2}}\left| 2 \right\rangle \left( \left| 1 \right\rangle -\left| 2 \right\rangle  \right), \\ 
		& \left| \psi _{3}^{\left( 1 \right)} \right\rangle =\frac{1}{\sqrt{2}}\left( \left| 1 \right\rangle +\left| 2 \right\rangle  \right)\left| 0 \right\rangle ,\quad \left| \psi _{3}^{\left( 2 \right)} \right\rangle =\frac{1}{\sqrt{2}}\left( \left| 1 \right\rangle -\left| 2 \right\rangle  \right)\left| 0 \right\rangle , \\
	\end{aligned}
\end{equation}

Let
\[\left| S \right\rangle =\frac{1}{3}\left( \left| 0 \right\rangle +\left| 1 \right\rangle +\left| 2 \right\rangle  \right)\left( \left| 0 \right\rangle +\left| 1 \right\rangle +\left| 2 \right\rangle  \right).\]

be a stopper state to force the unextendibility.

Then we claim that the set $\Psi $ is a UPB in ${{C}^{3}}\otimes {{C}^{3}}$ system. 

\[ \psi =\beta \cup \left\{ \left| S \right\rangle  \right\}\backslash \left\{ \left| \psi _{i}^{\left( 1 \right)} \right\rangle  \right\}_{i=1}^{3}. \]

The Shifts UPB in ${{C}^{2}}\otimes {{C}^{2}}\otimes {{C}^{2}}$ was proposed by Bennet in 1999, which provides one of the oldest examples of a nontrivial UPB. It consists of the following four states:

\begin{equation}
	\begin{aligned}
		& \left| {{\psi }_{1}} \right\rangle =\left| 0 \right\rangle \left| 1 \right\rangle \left| 0-1 \right\rangle , \\ 
		& \left| {{\psi }_{2}} \right\rangle =\left| 1 \right\rangle \left| 0-1 \right\rangle \left| 0 \right\rangle , \\ 
		& \left| {{\psi }_{3}} \right\rangle =\left| 0-1 \right\rangle \left| 0 \right\rangle \left| 1 \right\rangle , \\ 
		& \left| {{\psi }_{4}} \right\rangle =\left( \left| 0 \right\rangle +\left| 1 \right\rangle  \right)\left( \left| 0 \right\rangle +\left| 1 \right\rangle  \right)\left( \left| 0 \right\rangle +\left| 1 \right\rangle  \right). \\ 
	\end{aligned}
\end{equation}
	
This UPB can be simply generalized to a UPB over any number of parties, each with a one qubit Hilbert space. In this paper, the construction of a 3-qubit UPB is also based on the Shifts UPB. 

\section{Tripartite system with same dimensions}

Ref. \cite{31} proved that the members of UPB cannot be perfectly distinguishable by LOCC, that is, UPB is always nonlocal. Then, we set out to investigate what kind of multi-qubit UPB structure is strongly nonlocal. 

In Ref. \cite{43}, it was shown that starting from a two-qubit unextendible entangled basis, it is possible to construct a three-qubit unextendible entangled basis. Therefore, in this section, we proposed a set of strongly nonlocal UPB in ${{C}^{3}}\otimes {{C}^{3}}\otimes {{C}^{3}}$ based on Shifts UPB in Lemma 1. In the same way, we construct a set of UPB in ${{C}^{4}}\otimes {{C}^{4}}\otimes {{C}^{4}}$ in Lemma 2. Furthermore, we generalize these two constructions to ${{C}^{d}}\otimes {{C}^{d}}\otimes {{C}^{d}}$ system for any $d\ge 3$ in Theorem 1.

\subsection{Construct a UPB in ${{C}^{3}}\otimes {{C}^{3}}\otimes {{C}^{3}}$ system based on Shifts UPB}
Shifts UPB is one of the most classical available 3-qubit UPB. When the structure of it is shown by a Rubik's Cube, it can be found that any subset of two vectors on either side spans the two-dimensional space of that party, preventing any new vector from being orthogonal to all the existing ones. Following this idea, we construct a ${{C}^{3}}\otimes {{C}^{3}}\otimes {{C}^{3}}$ state set in Eq. (3).

\begin{equation}
	\begin{aligned}
	& {{C}_{1}}:\ \left\{ \begin{matrix}
		\left| {{\psi }_{0}} \right\rangle ={{\left| 1 \right\rangle }_{A}}{{\left| 0 \right\rangle }_{B}}{{\left| 0-1 \right\rangle }_{C}},  \\
		\left| {{\psi }_{1}} \right\rangle ={{\left| 0 \right\rangle }_{A}}{{\left| 0-1 \right\rangle }_{B}}{{\left| 1 \right\rangle }_{C}},  \\
		\left| {{\psi }_{2}} \right\rangle ={{\left| 0-1 \right\rangle }_{A}}{{\left| 1 \right\rangle }_{B}}{{\left| 0 \right\rangle }_{C}},  \\
	\end{matrix} \right. \\ 
	& {{C}_{2}}:\ \left\{ \begin{matrix}
		\left| {{\psi }_{3}} \right\rangle ={{\left| 1 \right\rangle }_{A}}{{\left| 2 \right\rangle }_{B}}{{\left| 1-2 \right\rangle }_{C}},  \\
		\left| {{\psi }_{4}} \right\rangle ={{\left| 2 \right\rangle }_{A}}{{\left| 1-2 \right\rangle }_{B}}{{\left| 1 \right\rangle }_{C}},  \\
		\left| {{\psi }_{5}} \right\rangle ={{\left| 1-2 \right\rangle }_{A}}{{\left| 1 \right\rangle }_{B}}{{\left| 2 \right\rangle }_{C}},  \\
	\end{matrix} \right. \\ 
	& {{C}_{3}}:\ \left\{ \begin{matrix}
		\left| {{\psi }_{6}} \right\rangle ={{\left| 2 \right\rangle }_{A}}{{\left| 0 \right\rangle }_{B}}{{\left| 1-2 \right\rangle }_{C}}  \\
		\left| {{\psi }_{7}} \right\rangle ={{\left| 1-2 \right\rangle }_{A}}{{\left| 2 \right\rangle }_{B}}{{\left| 0 \right\rangle }_{C}}  \\
		\left| {{\psi }_{8}} \right\rangle ={{\left| 0 \right\rangle }_{A}}{{\left| 1-2 \right\rangle }_{B}}{{\left| 2 \right\rangle }_{C}}  \\
		\left| {{\psi }_{9}} \right\rangle ={{\left| 0 \right\rangle }_{A}}{{\left| 2 \right\rangle }_{B}}{{\left| 0-1 \right\rangle }_{C}}  \\
		\left| {{\psi }_{10}} \right\rangle ={{\left| 0-1 \right\rangle }_{A}}{{\left| 0 \right\rangle }_{B}}{{\left| 2 \right\rangle }_{C}}  \\
		\left| {{\psi }_{11}} \right\rangle ={{\left| 2 \right\rangle }_{A}}{{\left| 0-1 \right\rangle }_{B}}{{\left| 0 \right\rangle }_{C}}  \\
	\end{matrix} \right. \\ 
	& Stopper:\ \left| {{\psi }_{12}} \right\rangle ={{\left( \left| 0 \right\rangle +\left| 1 \right\rangle +\left| 2 \right\rangle  \right)}_{A}}\otimes \\
	&{{\left( \left| 0 \right\rangle +\left| 1 \right\rangle +\left| 2 \right\rangle  \right)}_{B}}\otimes {{\left( \left| 0 \right\rangle +\left| 1 \right\rangle +\left| 2 \right\rangle  \right)}_{C}}. \\  
   \end{aligned}
\end{equation}

While $\left| {{\psi }_{12}} \right\rangle$ is considered as stopper states since it stops inclusion of any other states from forming a COPB. Geometrically, by mapping Eq. (3) to the Rubik's Cube, the structure of UPB can be displayed more intuitively and clearly, as shown in Fig. 2. The stopper state is not depicted. And we find that any two of these vectors are not in a two dimensional plane.

\begin{figure}[b]
	\includegraphics[width=0.35\textwidth]{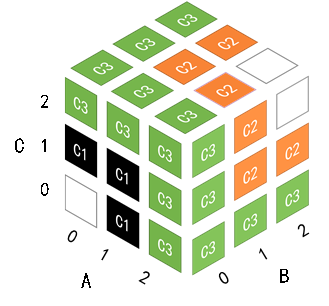}
	\caption{\label{fig:epsart} the Rubik's Cube corresponding to ${{C}^{3}}\otimes {{C}^{3}}\otimes {{C}^{3}}$.}
\end{figure}


The composition of UPB can be roughly divided into three parts:

C1: The set of UPB with $d=2$ in the lower left corner, which structure is similar to the Shifts UPB. It consists of three subcubes $\left\{ S{{C}_{1}},\ S{{C}_{2}},\ S{{C}_{3}} \right\}$, and is nonlocal, as shown in Figure 3-a.

C2: The set of UPB with $d=2$in the upper right corner, which structure is the same as the Shifts UPB. It includes three subcubes $\left\{ S{{C}_{1}},\ S{{C}_{2}},\ S{{C}_{3}} \right\}$, and is nonlocal, as shown in Figure 3-b.

C3: Six edges of the $3\times 3\times 3$ Rubik's Cube, corresponding to 6 subcubes $\left\{ S{{C}_{i}} \right\}_{i=1}^{6}$. This system is nonlocal, as shown in Fig. 3-c.

\begin{figure*}
	\includegraphics[width=0.75\textwidth]{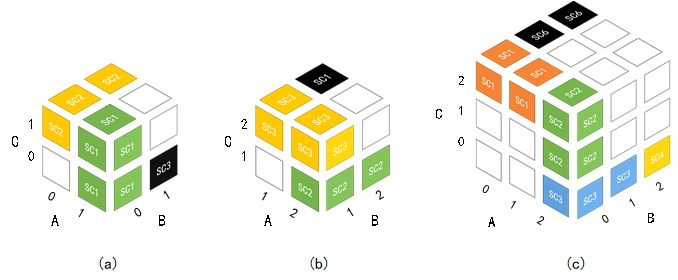}
	\caption{\label{fig:wide}composition of UPB in ${{C}^{3}}\otimes {{C}^{3}}\otimes {{C}^{3}}$.}
\end{figure*}

Each subsystem is nonlocal, so the whole system ${{C}_{1}}\cup {{C}_{2}}\cup {{C}_{3}}$ is also nonlocal. 

Considering two qubits in UPB, if they have the same orthogonal graph after permuting the qubits and relabeling the vertices, the two qubits are considered to be equivalent. Fig. 4-a is an original orthogonal graph of three qubits drawn respectively according to Eq. (5). The vertices ${{V}_{0}}\sim{{V}_{11}}$ in the graph correspond to $\left| {{\psi }_{0}} \right\rangle $ to $\left| {{\psi }_{11}} \right\rangle $ in the state set, and the lines between vertices represent the orthogonal relationship between two states. It can be seen that every edge between two vertices appears in at least one graph, indicating that all states in the set are mutually orthogonal. Fig. 4-b can be obtained after rearranging the vertices and corresponding lines. By observing Figure 4-b, we find that the three qubits \textit{A}, \textit{B}, and \textit{C} are equivalent because the orthogonal graph structure of each party is the same. In other words, by permuting each basis of ${{C}^{3}}$ used in the construction, the original UPB still can be obtained.

\begin{figure*}
	\includegraphics[width=0.85\textwidth]{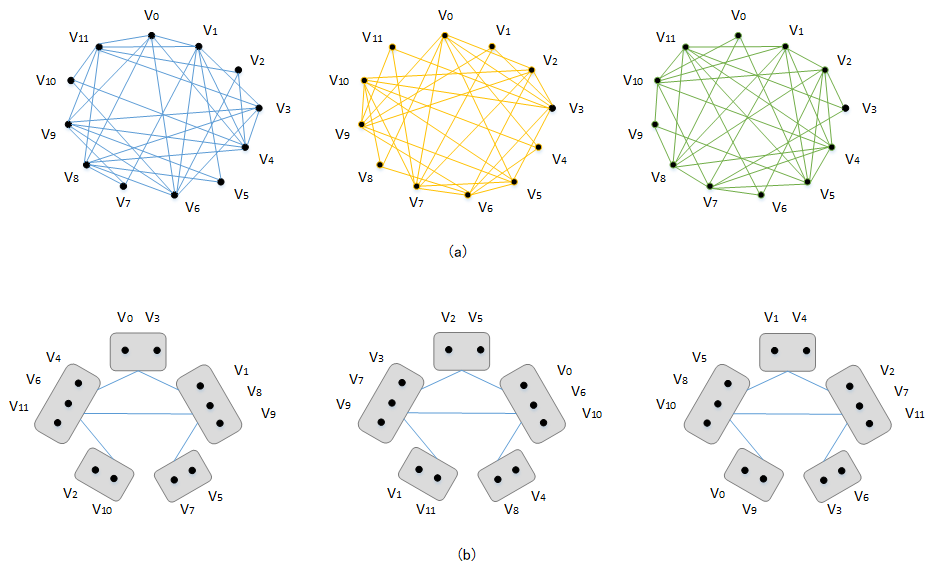}
	\caption{\label{fig:wide} Orthogonal graph of three qubits.}
\end{figure*}

Therefore, only the properties of one qubit need to be discussed,and then the properties of the other two qubits can be deduced from equivalence. From the composition of the basis, it is obvious that these states have a cyclic property as the cyclic property of the trace. In other words, the state set has the same properties in the different divisions of \textit{A|BC},\textit{B|AC} and \textit{C|AB}. In addition, the number 12 is also the minimum size that can be realized in an orthogonal product states set with strong nonlocality.

\textbf{Lemma 1 }: In ${{C}^{3}}\otimes {{C}^{3}}\otimes {{C}^{3}}$, the 3-qubit UPB of size 12 given by Eq.(3) is strongly nonlocal.

Proof: The proof can be summarized into two steps, and the first step is to prove that this 3-qubit system is nonlocal. This step can be omitted in the UPB set since UPB are incomplete and nonlocal.

The second step is to prove that the whole system remains nonlocal, regardless of the arbitrary partitioning of three qubits. Next, we use the division method of \textit{A|BC} as an example to carry out the proof.

Physically, this method of division means that the subsystems \textit{B}(Bob) and \textit{C}(Charlie) are treated together as a 9-dimensional subsystem \textit{BC} on which joint measurements are now allowed. On account of the original UPB is nonlocal, the system will be still nonlocal when Charlie goes first. Then, we only need to discuss the situation where the \textit{BC} system goes first. In order to make the proof process clearer, we first rewrite the original bases, let $\left| 00 \right\rangle \to \left| 0 \right\rangle ,\ \left| 01 \right\rangle \to \left| 1 \right\rangle ,\ \cdots ,\ \left| 23 \right\rangle \to \left| 8 \right\rangle $, and get the following states Eq. (4):

\begin{equation}
	\begin{aligned}
		& {{C}_{1}}:\ \left\{ \begin{matrix}
			\left| {{\psi }_{0}} \right\rangle ={{\left| 1 \right\rangle }_{A}}{{\left| 0-1 \right\rangle }_{BC}},  \\
			\left| {{\psi }_{1}} \right\rangle ={{\left| 0 \right\rangle }_{A}}{{\left| 1-4 \right\rangle }_{BC}},  \\
			\left| {{\psi }_{2}} \right\rangle ={{\left| 0-1 \right\rangle }_{A}}{{\left| 3 \right\rangle }_{BC}},  \\
		\end{matrix} \right. \\ 
		& {{C}_{2}}:\ \left\{ \begin{matrix}
			\left| {{\psi }_{3}} \right\rangle ={{\left| 1 \right\rangle }_{A}}{{\left| 7-8 \right\rangle }_{BC}},  \\
			\left| {{\psi }_{4}} \right\rangle ={{\left| 2 \right\rangle }_{A}}{{\left| 4-7 \right\rangle }_{BC}},  \\
			\left| {{\psi }_{5}} \right\rangle ={{\left| 1-2 \right\rangle }_{A}}{{\left| 5 \right\rangle }_{BC}},  \\
		\end{matrix} \right. \\ 
		& {{C}_{3}}:\,\ \left\{ \begin{matrix}
			\left| {{\psi }_{6}} \right\rangle ={{\left| 2 \right\rangle }_{A}}{{\left| 1-2 \right\rangle }_{BC}}  \\
			\left| {{\psi }_{7}} \right\rangle ={{\left| 1-2 \right\rangle }_{A}}{{\left| 6 \right\rangle }_{BC}}  \\
			\left| {{\psi }_{8}} \right\rangle ={{\left| 0 \right\rangle }_{A}}{{\left| 5-8 \right\rangle }_{BC}}  \\
			\left| {{\psi }_{9}} \right\rangle ={{\left| 0 \right\rangle }_{A}}{{\left| 6-7 \right\rangle }_{BC}}  \\
			\left| {{\psi }_{10}} \right\rangle ={{\left| 0-1 \right\rangle }_{A}}{{\left| 2 \right\rangle }_{BC}}  \\
			\left| {{\psi }_{11}} \right\rangle ={{\left| 2 \right\rangle }_{A}}{{\left| 0-3 \right\rangle }_{BC}}  \\
		\end{matrix} \right. \\ 
		& Stopper:\ \left| {{\psi }_{12}} \right\rangle ={{\left( \left| 0 \right\rangle +\cdots +\left| 8 \right\rangle  \right)}_{A}}{{\left( \left| 0 \right\rangle +\left| 1 \right\rangle +\left| 2 \right\rangle  \right)}_{BC}}. \\  
   \end{aligned}
\end{equation}

The 12 states in \textit{A|BC} bipartition correspond to 12 blocks of the $3\times 9$ grid in Fig. 5. The yellow grid represents ${{C}_{1}}$ part, the blue grid represents ${{C}_{2}}$ part, and the green grid represents ${{C}_{3}}$ part. The whole graph is centrosymmetric.

\begin{figure*}
	\includegraphics[width=0.7\textwidth]{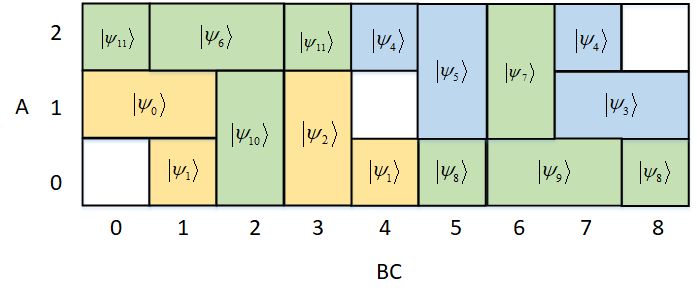}
	\caption{\label{fig:wide}The corresponding 3 × 9 grid of Eq. (4).}
\end{figure*}

Suppose \textit{BC} system starts with the nontrivial and non-disturbing measurement, represented by a set of POVM elements $M_{m}^{\dagger }M_{m}^{{}}$ on ${{d}^{2}}\times {{d}^{2}}$. the POVM measurement in ${{\left\{ \left| 0 \right\rangle ,\left| 1 \right\rangle ,\cdots ,\left| 8 \right\rangle  \right\}}_{A}}$ basis can be written, which corresponds to the states Eq. (4):

Therefore, the original matrix can be reduced to:

\begin{eqnarray*}
M_{m}^{\dagger }M_{m}^{{}}=\left[ \begin{matrix}
	{{a}_{00}} & {{a}_{01}} & \cdots  & {{a}_{07}} & {{a}_{08}}  \\
	{{a}_{10}} & {{a}_{11}} & \cdots  & {{a}_{17}} & {{a}_{18}}  \\
	\vdots  & \vdots  & \ddots  & \vdots  & \vdots   \\
	{{a}_{70}} & {{a}_{71}} & \cdots  & {{a}_{77}} & {{a}_{78}}  \\
	{{a}_{80}} & {{a}_{81}} & \cdots  & {{a}_{87}} & {{a}_{88}}  \\
\end{matrix} \right]
\end{eqnarray*}

The post-measurement states could be expressed as $\left( I\otimes {{M}_{m}} \right)\left| {{\varphi }_{i}} \right\rangle $, which should be mutually orthogonal. Then $\left\langle  {{\varphi }_{j}} \right|\left( I\otimes M_{m}^{\dagger }M_{m}^{{}} \right)\left| {{\varphi }_{i}} \right\rangle =0$ is obtained. According to this principle, the original matrix could be transformed into:

\begin{eqnarray*}
M_{m}^{\dagger }M_{m}^{{}}=\left[ \begin{matrix}
	a & 0 & \cdots  & 0  \\
	0 & a & \cdots  & 0  \\
	\vdots  & \vdots  & \ddots  & \vdots   \\
	0 & 0 & \cdots  & {{a}_{{}}}  \\
\end{matrix} \right]
\end{eqnarray*}

Table I shows the detailed derivation process.
\begin{table*}
	\caption{\label{tab:tableA1} POVM elements}
	\begin{ruledtabular}
		\begin{tabular}{cccccc}
			POVM Element & \multicolumn{5}{c}{Corresponding States} \\
			\hline
			\multirow{4}{*}{${{a}_{0i}}={{a}_{i0}}=0$} & $i=1$ & $i=2$ & $i=3$ & $i=4$ & $i=5$\\
			&$\left| {{\varphi }_{11}} \right\rangle $,$\left| {{\varphi }_{6}} \right\rangle $ &$\left| {{\varphi }_{11}} \right\rangle $,$\left| {{\varphi }_{10}} \right\rangle $ &$\left| {{\varphi }_{0}} \right\rangle $,$\left| {{\varphi }_{2}} \right\rangle $ &$\left| {{\varphi }_{0}} \right\rangle $,$\left| {{\varphi }_{4}} \right\rangle $ &$\left| {{\varphi }_{0}} \right\rangle $,$\left| {{\varphi }_{5}} \right\rangle $ \\
			&$i=6$ &$i=7$ &$i=8$ & &\\
			&$\left| {{\varphi }_{0}} \right\rangle $,$\left| {{\varphi }_{7}} \right\rangle $ &$\left| {{\varphi }_{0}} \right\rangle $,$\left| {{\varphi }_{9}} \right\rangle $ &$\left| {{\varphi }_{0}} \right\rangle $,$\left| {{\varphi }_{3}} \right\rangle $ & &\\
			\hline
			\multirow{4}{*}{${{a}_{1i}}={{a}_{i1}}=0$} & $i=2$ & $i=3$ & $i=4$ & $i=5$ &$i=6$\\
			&$\left| {{\varphi }_{0}} \right\rangle $,$\left| {{\varphi }_{10}} \right\rangle $ &$\left| {{\varphi }_{6}} \right\rangle $,$\left| {{\varphi }_{2}} \right\rangle $ &$\left| {{\varphi }_{6}} \right\rangle $,$\left| {{\varphi }_{4}} \right\rangle $ &$\left| {{\varphi }_{6}} \right\rangle $,$\left| {{\varphi }_{5}} \right\rangle $ &$\left| {{\varphi }_{6}} \right\rangle $,$\left| {{\varphi }_{7}} \right\rangle $\\
			&$i=7$ &$i=8$ & & & \\
			&$\left| {{\varphi }_{6}} \right\rangle $,$\left| {{\varphi }_{9}} \right\rangle $ &$\left| {{\varphi }_{6}} \right\rangle $,$\left| {{\varphi }_{3}} \right\rangle $ & & & \\
			\hline
			\multirow{4}{*}{${{a}_{2i}}={{a}_{i2}}=0$} & $i=3$ & $i=4$ & $i=5$ & $i=6$ & $i=7$\\
			&$\left| {{\varphi }_{10}} \right\rangle $,$\left| {{\varphi }_{2}} \right\rangle $ &$\left| {{\varphi }_{10}} \right\rangle $,$\left| {{\varphi }_{4}} \right\rangle $ &$\left| {{\varphi }_{10}} \right\rangle $,$\left| {{\varphi }_{5}} \right\rangle $ &$\left| {{\varphi }_{10}} \right\rangle $,$\left| {{\varphi }_{7}} \right\rangle $ &$\left| {{\varphi }_{10}} \right\rangle $,$\left| {{\varphi }_{9}} \right\rangle $ \\
			& $i=8$ & & & & \\
			&$\left| {{\varphi }_{10}} \right\rangle $,$\left| {{\varphi }_{3}} \right\rangle $ & & & & \\
			\hline
			\multirow{2}{*}{${{a}_{3i}}={{a}_{i3}}=0$} & $i=4$ & $i=5$ & $i=6$ & $i=7$& $i=8$ \\
			&$\left| {{\varphi }_{2}} \right\rangle $,$\left| {{\varphi }_{4}} \right\rangle $ &$\left| {{\varphi }_{2}} \right\rangle $,$\left| {{\varphi }_{5}} \right\rangle $ 
			&$\left| {{\varphi }_{2}} \right\rangle $,$\left| {{\varphi }_{7}} \right\rangle $ &$\left| {{\varphi }_{2}} \right\rangle $,$\left| {{\varphi }_{9}} \right\rangle $ &$\left| {{\varphi }_{2}} \right\rangle $,$\left| {{\varphi }_{3}} \right\rangle $ \\
			\hline
			\multirow{2}{*}{${{a}_{4i}}={{a}_{i4}}=0$} & $i=5$ & $i=6$ & $i=7$& $i=8$ & \\
			&$\left| {{\varphi }_{1}} \right\rangle $,$\left| {{\varphi }_{5}} \right\rangle $ &$\left| {{\varphi }_{1}} \right\rangle $,$\left| {{\varphi }_{7}} \right\rangle $ 
			&$\left| {{\varphi }_{1}} \right\rangle $,$\left| {{\varphi }_{9}} \right\rangle $ &$\left| {{\varphi }_{1}} \right\rangle $,$\left| {{\varphi }_{3}} \right\rangle $ &\\
			\hline
			\multirow{2}{*}{${{a}_{5i}}={{a}_{i5}}=0$} & $i=6$ & $i=7$& $i=8$ & &  \\
			&$\left| {{\varphi }_{5}} \right\rangle $,$\left| {{\varphi }_{7}} \right\rangle $ &$\left| {{\varphi }_{5}} \right\rangle $,$\left| {{\varphi }_{9}} \right\rangle $ &$\left| {{\varphi }_{5}} \right\rangle $,$\left| {{\varphi }_{3}} \right\rangle $ &  &  \\
			\hline
			\multirow{2}{*}{${{a}_{6i}}={{a}_{i6}}=0$} & $i=7$& $i=8$ &  &  &  \\
			&$\left| {{\varphi }_{7}} \right\rangle $,$\left| {{\varphi }_{4}} \right\rangle $ &$\left| {{\varphi }_{7}} \right\rangle $,$\left| {{\varphi }_{8}} \right\rangle $ & &  & \\
			\hline
			\multirow{2}{*}{${{a}_{7i}}={{a}_{i7}}=0$} &$i=8$ &  &  & &  \\
			&$\left| {{\varphi }_{4}} \right\rangle $,$\left| {{\varphi }_{8}} \right\rangle$ & & & & \\
			\hline
			\multirow{4}{*}{${{a}_{00}}={{a}_{ii}}$} & $i=1$ & $i=2$ & $i=3$ & $i=4$ & $i=5$ \\
			&$\left| {{\varphi }_{0,12}} \right\rangle $ &$\left| {{\varphi }_{6,12}} \right\rangle $ &$\left| {{\varphi }_{11,12}} \right\rangle $ &$\left| {{\varphi }_{1,12}} \right\rangle $ &$\left| {{\varphi }_{8,12}} \right\rangle $\\
			& $i=6$ & $i=7$& $i=8$ &  & \\
			&$\left| {{\varphi }_{9,12}} \right\rangle $ &$\left| {{\varphi }_{4,12}} \right\rangle $ &$\left| {{\varphi }_{3,12}} \right\rangle $ & & \\
		\end{tabular}
	\end{ruledtabular}
\end{table*}

Obviously, \textit{BC}’s measurement matrix is proportional to the identity matrix, so it means that \textit{BC} system starts with a trivial measurement, and cannot get any information from the measurement result. As for the other two division methods, \textit{AB|C}, \textit{AC|B}, the proof method is similar to this. In summary, the strong nonlocality of UPB can be maintained by arbitrarily dividing the three qubits into two parts.

\subsection{Construct a UPB in ${{C}^{4}}\otimes {{C}^{4}}\otimes {{C}^{4}}$ system based on a UPB in ${{C}^{3}}\otimes {{C}^{3}}\otimes {{C}^{3}}$ system}

Based on the construction method of state set Eq. (3), we extend the set to the case of $d=4$. Compared with the Rubik's Cube of $3\times 3\times 3$, the cube of $4\times 4\times 4$ has three new planes, $\left\{ \left( 0,\ \cdots ,\ 3 \right),\ \left( 0,\ \cdots ,\ 3 \right),\ 3 \right\}$, $\left\{ \left( 0,\ \cdots ,\ 3 \right),\ 3,\ \left( 0,\ \cdots ,\ 3 \right) \right\}$, $\left\{ 3,\ \left( 0,\ \cdots ,\ 3 \right),\ \left( 0,\ \cdots ,\ 3 \right) \right\}$. According to tile structure, we divide the newly added plane and obtain Fig. 6.

\begin{figure}[b]
	\includegraphics[width=0.35\textwidth]{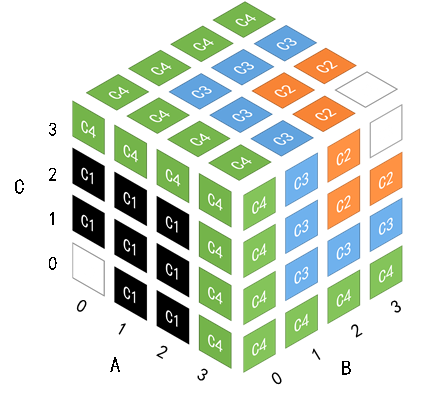}
	\caption{\label{fig:epsart} composition of UPB in ${{C}^{4}}\otimes {{C}^{4}}\otimes {{C}^{4}}$.}
\end{figure}


The UPB of $d=4$ can be roughly divided into the following four parts:

C1: The UPB of ${{C}^{3}}\otimes {{C}^{3}}\otimes {{C}^{3}}$ in the lower left corner. It includes 12 subcubes and is nonlocal, as shown in Figure 7-a.

C2: The set of UPB with $d=2$in the upper right corner, which structure is the same as the Shifts UPB. It includes three subcubes and is nonlocal, as shown in Figure 7-b.

C3: The remaining parts of the three newly added planes. It can be decomposed into six subcubes that cannot be further expanded, which is equivalent to shifting all the previous C3 parts (six edges in the system $d=3$) up and back by one grid. This system is non-local, as shown in Figure 7-c.

C4: Six edges of the $4\times 4\times 4$ Rubik's Cube, corresponding to 6 subcubes. This system is nonlocal, as shown in Fig. 7-d.

\begin{figure*}
	\includegraphics[width=0.75\textwidth]{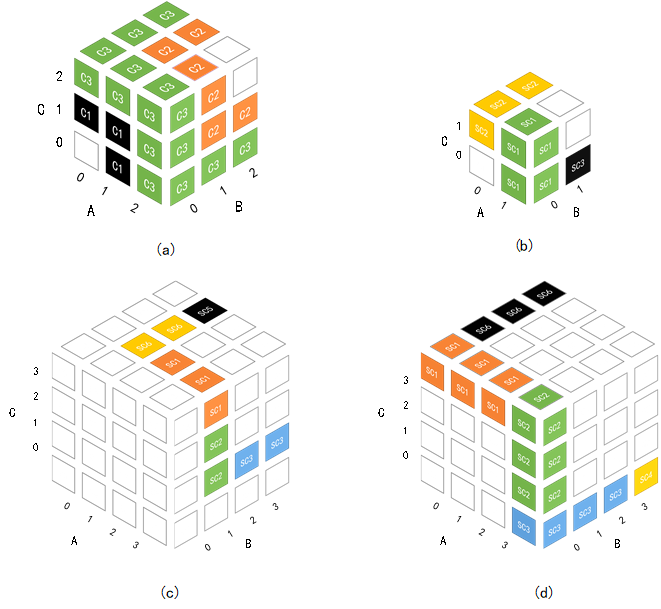}
	\caption{\label{fig:wide}composition of UPB in ${{C}^{4}}\otimes {{C}^{4}}\otimes {{C}^{4}}$.}
\end{figure*}

By matching the Rubik's cube with the states set, we get the Eq. (5). Compared with ${{C}^{3}}\otimes {{C}^{3}}\otimes {{C}^{3}}$, there is one more part, ${{C}_{3}}$ in the composition of Eq. (5), which is the basis of the newly added plane that has not been decomposed. 
\begin{subequations}
\begin{equation}
	\begin{aligned}
		& {{C}_{1}}:\ \left\{ \begin{matrix}
			\left| {{\psi }_{0}} \right\rangle ={{\left| 1 \right\rangle }_{A}}{{\left| 0 \right\rangle }_{B}}{{\left| 0-1 \right\rangle }_{C}},  \\
			\left| {{\psi }_{1}} \right\rangle ={{\left| 0 \right\rangle }_{A}}{{\left| 0-1 \right\rangle }_{B}}{{\left| 1 \right\rangle }_{C}},  \\
			\left| {{\psi }_{2}} \right\rangle ={{\left| 0-1 \right\rangle }_{A}}{{\left| 1 \right\rangle }_{B}}{{\left| 0 \right\rangle }_{C}},  \\
			\left| {{\psi }_{3}} \right\rangle ={{\left| 1 \right\rangle }_{A}}{{\left| 2 \right\rangle }_{B}}{{\left| 1-2 \right\rangle }_{C}},  \\
			\left| {{\psi }_{4}} \right\rangle ={{\left| 2 \right\rangle }_{A}}{{\left| 1-2 \right\rangle }_{B}}{{\left| 1 \right\rangle }_{C}},  \\
			\left| {{\psi }_{5}} \right\rangle ={{\left| 1-2 \right\rangle }_{A}}{{\left| 1 \right\rangle }_{B}}{{\left| 2 \right\rangle }_{C}},  \\
			\left| {{\psi }_{6}} \right\rangle ={{\left| 2 \right\rangle }_{A}}{{\left| 0 \right\rangle }_{B}}{{\left| 1-2 \right\rangle }_{C}},  \\
			\left| {{\psi }_{7}} \right\rangle ={{\left| 1-2 \right\rangle }_{A}}{{\left| 2 \right\rangle }_{B}}{{\left| 0 \right\rangle }_{C}},  \\
			\left| {{\psi }_{8}} \right\rangle ={{\left| 0 \right\rangle }_{A}}{{\left| 1-2 \right\rangle }_{B}}{{\left| 2 \right\rangle }_{C}},  \\
			\left| {{\psi }_{9}} \right\rangle ={{\left| 0 \right\rangle }_{A}}{{\left| 2 \right\rangle }_{B}}{{\left| 0-1 \right\rangle }_{C}},  \\
			\left| {{\psi }_{10}} \right\rangle ={{\left| 0-1 \right\rangle }_{A}}{{\left| 0 \right\rangle }_{B}}{{\left| 2 \right\rangle }_{C}},  \\
			\left| {{\psi }_{11}} \right\rangle ={{\left| 2 \right\rangle }_{A}}{{\left| 0-1 \right\rangle }_{B}}{{\left| 0 \right\rangle }_{C}},  \\
		\end{matrix} \right. \\ 
		& {{C}_{2}}:\ \left\{ \begin{matrix}
			\left| {{\psi }_{12}} \right\rangle ={{\left| 3 \right\rangle }_{A}}{{\left| 2 \right\rangle }_{B}}{{\left| 2-3 \right\rangle }_{C}},  \\
			\left| {{\psi }_{13}} \right\rangle ={{\left| 2 \right\rangle }_{A}}{{\left| 2-3 \right\rangle }_{B}}{{\left| 3 \right\rangle }_{C}},  \\
			\left| {{\psi }_{14}} \right\rangle ={{\left| 2-3 \right\rangle }_{A}}{{\left| 3 \right\rangle }_{B}}{{\left| 2 \right\rangle }_{C}},  \\
		\end{matrix} \right. \\ 
		& {{C}_{3}}:\ \left\{ \begin{matrix}
			\left| {{\psi }_{15}} \right\rangle ={{\left| 1-2 \right\rangle }_{A}}{{\left| 3 \right\rangle }_{B}}{{\left| 1 \right\rangle }_{C}},  \\
			\left| {{\psi }_{16}} \right\rangle ={{\left| 1 \right\rangle }_{A}}{{\left| 1-2 \right\rangle }_{B}}{{\left| 3 \right\rangle }_{C}},  \\
			\left| {{\psi }_{17}} \right\rangle ={{\left| 3 \right\rangle }_{A}}{{\left| 1 \right\rangle }_{B}}{{\left| 1-2 \right\rangle }_{C}},  \\
			\left| {{\psi }_{18}} \right\rangle ={{\left| 2-3 \right\rangle }_{A}}{{\left| 1 \right\rangle }_{B}}{{\left| 3 \right\rangle }_{C}},  \\
			\left| {{\psi }_{19}} \right\rangle ={{\left| 3 \right\rangle }_{A}}{{\left| 2-3 \right\rangle }_{B}}{{\left| 1 \right\rangle }_{C}},  \\
			\left| {{\psi }_{20}} \right\rangle ={{\left| 1 \right\rangle }_{A}}{{\left| 3 \right\rangle }_{B}}{{\left| 2-3 \right\rangle }_{C}},  \\
		\end{matrix} \right. \\ 
\end{aligned}
\end{equation}
\begin{equation}
\begin{aligned}
		& {{C}_{4}}:\ \left\{ \begin{array}{*{35}{l}}
			\left| {{\psi }_{21,22}} \right\rangle ={{\left| 0 \right\rangle }_{A}}{{\left| 3 \right\rangle }_{B}}{{\left| 0+w_{4}^{s}1+w_{4}^{2s}2 \right\rangle }_{C}},  \\
			\left| {{\psi }_{23.24}} \right\rangle ={{\left| 0+w_{4}^{s}1+w_{4}^{2s}2 \right\rangle }_{A}}{{\left| 0 \right\rangle }_{B}}{{\left| 3 \right\rangle }_{C}},  \\
			\left| {{\psi }_{25,26}} \right\rangle ={{\left| 3 \right\rangle }_{A}}{{\left| 0+w_{4}^{s}1+w_{4}^{2s}2 \right\rangle }_{B}}{{\left| 0 \right\rangle }_{C}},  \\
			\left| {{\psi }_{27,28}} \right\rangle ={{\left| 3 \right\rangle }_{A}}{{\left| 0 \right\rangle }_{B}}{{\left| 1+w_{4}^{s}2+w_{4}^{2s}3 \right\rangle }_{C}},  \\
			\left| {{\psi }_{29.30}} \right\rangle ={{\left| 1+w_{4}^{s}2+w_{4}^{2s}3 \right\rangle }_{A}}{{\left| 3 \right\rangle }_{B}}{{\left| 0 \right\rangle }_{C}},  \\
			\left| {{\psi }_{31,32}} \right\rangle ={{\left| 0 \right\rangle }_{A}}{{\left| 1+w_{4}^{s}2+w_{4}^{2s}3 \right\rangle }_{B}}{{\left| 3 \right\rangle }_{C}},  \\
		\end{array} \right. \\ 
   		& Stopper:\ \left| {{\psi }_{30}} \right\rangle ={{\left( \left| 0 \right\rangle +\left| 1 \right\rangle +\left| 2 \right\rangle +\left| 3 \right\rangle  \right)}_{A}}\otimes \\
		& {{\left( \left| 0 \right\rangle +\left| 1 \right\rangle +\left| 2 \right\rangle +\left| 3 \right\rangle  \right)}_{B}}\otimes {{\left( \left| 0 \right\rangle +\left| 1 \right\rangle +\left| 2 \right\rangle +\left| 3 \right\rangle  \right)}_{C}}. \\ 	 	 
	 \end{aligned}
\end{equation}
\end{subequations}

\textbf{Lemma 2: } In ${{C}^{4}}\otimes {{C}^{4}}\otimes {{C}^{4}}$, the 3-qubit UPB of size 34 given by Eq. (5) is strongly nonlocal. 

The proof of Lemma 2 is given in Appendix A.

\subsection{Construct a UPB in ${{C}^{d}}\otimes {{C}^{d}}\otimes {{C}^{d}}$ system based on a UPB in ${{C}^{d-1}}\otimes {{C}^{d-1}}\otimes {{C}^{d-1}}$ system}

After analyzing the UPB structure constructed in Lemma1 and Lemma2, we propose a UPB in ${{C}^{d}}\otimes {{C}^{d}}\otimes {{C}^{d}}$ of size ${{\left( d-1 \right)}^{3}}+3\left( d-2 \right)+1$ based on the set ${{C}^{d-1}}\otimes {{C}^{d-1}}\otimes {{C}^{d-1}}$, and then obtain Proposition 1.

The composition of the general 3-qubit UPB is closely related to the composition of the Rubik's Cube and is always built on the basis of lower dimensional state set. Thus, the division of a Rubik's cube can also be  based on low-dimensional cubes, like peeling an onion. First, the three outer layers, $\left\{ \left( 0,\ \cdots ,\ d-1 \right),\ \left( 0,\ \cdots ,\ d-1 \right),\ d-1 \right\}$, $\left\{ \left( 0,\ \cdots ,\ d-1 \right),\ d-1,\ \left( 0,\ \cdots ,\ d-1 \right) \right\}$, $\left\{ d-1,\ \left( 0,\ \cdots ,\ d-1 \right),\ \left( 0,\ \cdots ,\ d-1 \right) \right\}$ are divided into different nonlocal blocks. Then, the inside cube is a tripartite system with $d=d-1$.  Again, we divide the three outer layers, $\left\{ \left( 0,\ \cdots ,\ d-2 \right),\ \left( 0,\ \cdots ,\ d-2 \right),\ d-2 \right\}$, $\left\{ \left( 0,\ \cdots ,\ d-2 \right),\ d-2,\ \left( 0,\ \cdots ,\ d-2 \right) \right\}$, $\left\{ d-2,\ \left( 0,\ \cdots ,\ d-2 \right),\ \left( 0,\ \cdots ,\ d-2 \right) \right\}$ in the same way. And so on, until finally a 3-qubit UPB with $d=2$ is left, we will divide it according to the Shifts UPB.

\begin{figure*}
	\includegraphics[width=0.5\textwidth]{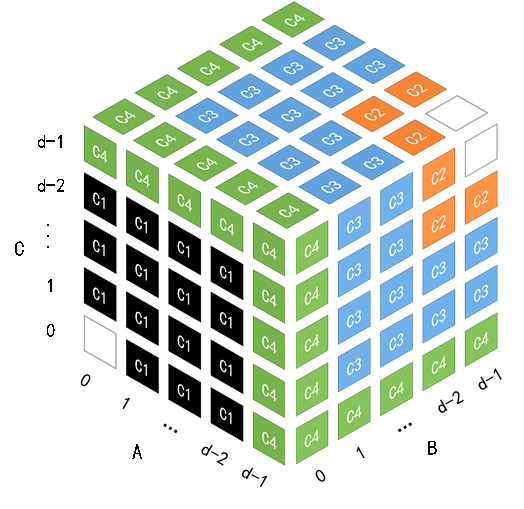}
	\caption{\label{fig:wide}composition of UPB in ${{C}^{d}}\otimes {{C}^{d}}\otimes {{C}^{d}}$.}
\end{figure*}

\textbf{Proposion 1: }In ${{C}^{d}}\otimes {{C}^{d}}\otimes {{C}^{d}}$, the UPB of size ${{\left( d-1 \right)}^{3}}+3\left( d-2 \right)+1$ given by Eq. (6) is strongly nonlocal.
\begin{widetext}
\begin{equation}
\begin{aligned}	
	& {{C}_{1}}:\ C_{1}^{d-1}\otimes C_{1}^{d-1}\otimes C_{1}^{d-1} \\ 
	& {{C}_{2}}:\ C_{2}^{d-1}+1 \\ 
	& {{C}_{3}}:\ C_{3}^{d-1}+1,\quad C_{4}^{d-1}+1 \\ 
	& {{C}_{4}}:\ \left\{ \begin{array}{*{35}{l}}
		\left| {{\psi }_{{{\left( d-1 \right)}^{3}}-3\left( d-2 \right)}} \right\rangle ={{\left| 0+\cdots +w_{d}^{\left( d-2 \right)s}\left( d-2 \right) \right\rangle }_{A}}{{\left| 0 \right\rangle }_{B}}{{\left| i \right\rangle }_{C}}  \\
		\left| {{\psi }_{{{\left( d-1 \right)}^{3}}-2\left( d-2 \right)}} \right\rangle ={{\left| i \right\rangle }_{A}}{{\left| 0+\cdots +w_{d}^{\left( d-2 \right)s}\left( d-2 \right) \right\rangle }_{B}}{{\left| 0 \right\rangle }_{C}}  \\
		\left| {{\psi }_{{{\left( d-1 \right)}^{3}}-\left( d-2 \right)}} \right\rangle ={{\left| 0 \right\rangle }_{A}}{{\left| i \right\rangle }_{B}}{{\left| 0+\cdots +w_{d}^{\left( d-2 \right)s}\left( d-2 \right) \right\rangle }_{C}}  \\
		\left| {{\psi }_{{{\left( d-1 \right)}^{3}}}} \right\rangle ={{\left| 1+\cdots +w_{d}^{\left( d-1 \right)s}\left( d-1 \right) \right\rangle }_{A}}{{\left| i \right\rangle }_{B}}{{\left| 0 \right\rangle }_{C}}  \\
		\left| {{\psi }_{{{\left( d-1 \right)}^{3}}+\left( d-2 \right)}} \right\rangle ={{\left| 0 \right\rangle }_{A}}{{\left| 1+\cdots +w_{d}^{\left( d-1 \right)s}\left( d-1 \right) \right\rangle }_{B}}{{\left| i \right\rangle }_{C}}  \\
		\left| {{\psi }_{{{\left( d-1 \right)}^{3}}+2\left( d-2 \right)}} \right\rangle ={{\left| i \right\rangle }_{A}}{{\left| 0 \right\rangle }_{B}}{{\left| 1+\cdots +w_{d}^{\left( d-1 \right)s}\left( d-1 \right) \right\rangle }_{C}}  \\
	\end{array} \right. \\ 
	& Stopper:\ \left| {{\psi }_{{{\left( d-1 \right)}^{3}}+3\left( d-2 \right)}} \right\rangle ={{\left( \left| 0 \right\rangle +\cdots +\left| d-1 \right\rangle  \right)}_{A}}{{\left( \left| 0 \right\rangle +\cdots +\left| d-1 \right\rangle  \right)}_{B}}{{\left( \left| 0 \right\rangle +\cdots +\left| d-1 \right\rangle  \right)}_{C}}. \\
\end{aligned}	
\end{equation}
\end{widetext}

Where $C_{k}^{d-1}\otimes C_{k}^{d-1}\otimes C_{k}^{d-1},\ k=1,2,3,4$ represents the ${{C}_{k}}$ part of the ${{C}^{d-1}}\otimes {{C}^{d-1}}\otimes {{C}^{d-1}}$ system. And  $C_{3}^{d-1}+1$ means the simultaneous incrementing of three parties in the $C_{3}^{d-1}\otimes C_{3}^{d-1}\otimes C_{3}^{d-1}$ system. 

The proposed UPB is according to:

\[\Psi =\beta \cup \left\{ \left| S \right\rangle  \right\}\backslash \left\{ \left| \psi _{i}^{\left( 0,\ 0,\ 0 \right)} \right\rangle  \right\}_{i=0}^{{}}\]

Among them, the subtracted set $\left\{ \left| \psi _{i}^{\left( 0,\ 0,\ 0 \right)} \right\rangle  \right\}_{i=0}^{2}$ is called the missing states, which are not orthogonal to $\left| S \right\rangle $ but orthogonal to all states in $\Psi \backslash \left\{ \left| S \right\rangle  \right\}$. So any state in $H_{\psi }^{\bot }$ must be a linear combination of the missing states. It is valuable in the discussion of bound entangled states.

The UPB of ${{C}^{d}}\otimes {{C}^{d}}\otimes {{C}^{d}}$ can be roughly divided into the following four parts:

C1: The set of UPB with $d=d-1$ in the lower left corner. It includes 12 subcubes and is nonlocal, as shown in Figure 9-a.

C2: The set of UPB with $d=2$ in the upper right corner, which structure is the same as the Shifts UPB. It includes three subcubes and is nonlocal, as shown in Figure 9-b.

C3: The remaining parts of the three newly added planes. It can be decomposed into six subcubes that cannot be further expanded, which is equivalent to shifting all the previous C3 and C4 parts of ${{C}^{d-1}}\otimes {{C}^{d-1}}\otimes {{C}^{d-1}}$ system, up and back by one grid. This system is non-local, as shown in Figure 9-c.

C4: Six edges of the $d\times d\times d$ Rubik's Cube, corresponding to 6 subcubes. This system is nonlocal, as shown in Fig. 9-d.

\begin{figure*}
	\includegraphics[width=0.75\textwidth]{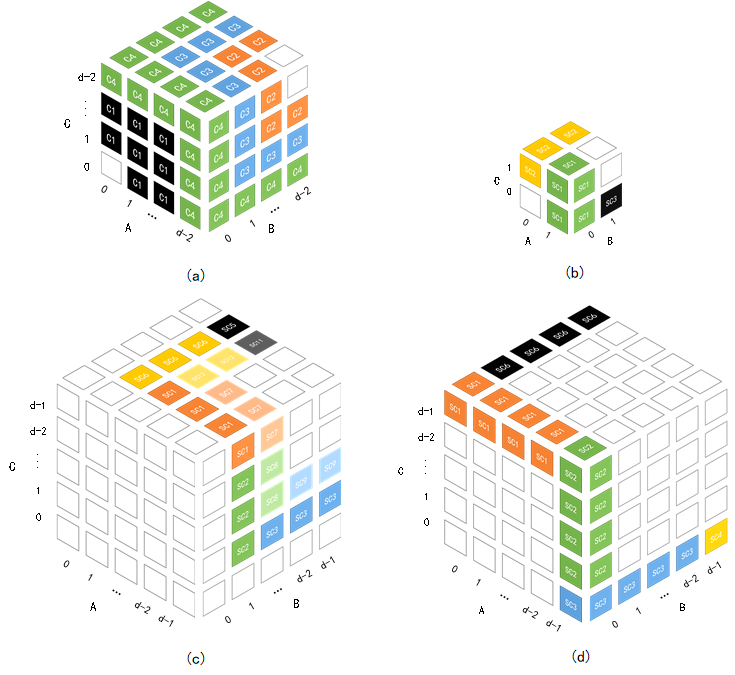}
	\caption{\label{fig:wide}composition of UPB in ${{C}^{d}}\otimes {{C}^{d}}\otimes {{C}^{d}}$.}
\end{figure*}

\section{\label{sec:level2}Tripartite system with different dimensions}

In the previous section, we discussed the UPB with the same dimensions of three parties. In this section, we continue to discuss another scenario which is more general in tripartite systems and propose a set of strongly nonlocal UPB in ${{C}^{{{d}_{1}}}}\otimes {{C}^{{{d}_{2}}}}\otimes {{C}^{{{d}_{3}}}},\quad d{}_{1},\ d{}_{2},\ d{}_{3}\,\in \left[ 0,\ d-1 \right]$.

\subsection{A Construct a UPB in ${{C}^{3}}\otimes {{C}^{3}}\otimes {{C}^{4}}$ system with strong nonlocality}
Tile structure is a classical approach in the general construction of 2-qubit UPB, which was first proposed by Bennet in 1999. Later, many researchers studied it and proposed some related structures, such as Gen tile structure, etc. In tile structure, any two sub-rectangles (i.e. tiles) cannot be combined to form a new rectangle by a simple translation. In other words, any rectangle cannot be split into two smaller sub-rectangles. Following this method, we consider whether tile structure can be extended to 3-qubit.

The tripartite system ${{C}^{{{d}_{1}}}}\otimes {{C}^{{{d}_{2}}}}\otimes {{C}^{{{d}_{3}}}}$ can always be uniquely mapped to a Rubik's Cube whose section is one of these three planes, ${{C}^{{{d}_{1}}}}\otimes {{C}^{{{d}_{2}}}}$, ${{C}^{{{d}_{2}}}}\otimes {{C}^{{{d}_{3}}}}$ and ${{C}^{{{d}_{1}}}}\otimes {{C}^{{{d}_{3}}}}$. Therefore, we extend the classical tile structure to three parties, which is defined as follows:

\textbf{Definition 5: } If two sub-cuboids (i.e. tiles) cannot be combined to form a new cuboid (i.e. tiles) by a simple translation, then the structure of the state set is defined as Tri-tile structure. In other words, any two vectors are not in a two dimensional plane.

Moreover, we observe that the state set proposed in the previous section also satisfies the Tri-tile structure. Take $d=3$ as an example, Fig. 10 can be obtained, which displays all 9 sections of the cube. It can be seen that these sections all meet the requirements of tile structure, so this ${{C}^{3}}\otimes {{C}^{3}}\otimes {{C}^{3}}$ system satisfies Tri-tile structure. 

\begin{figure*}
	\includegraphics[width=0.75\textwidth]{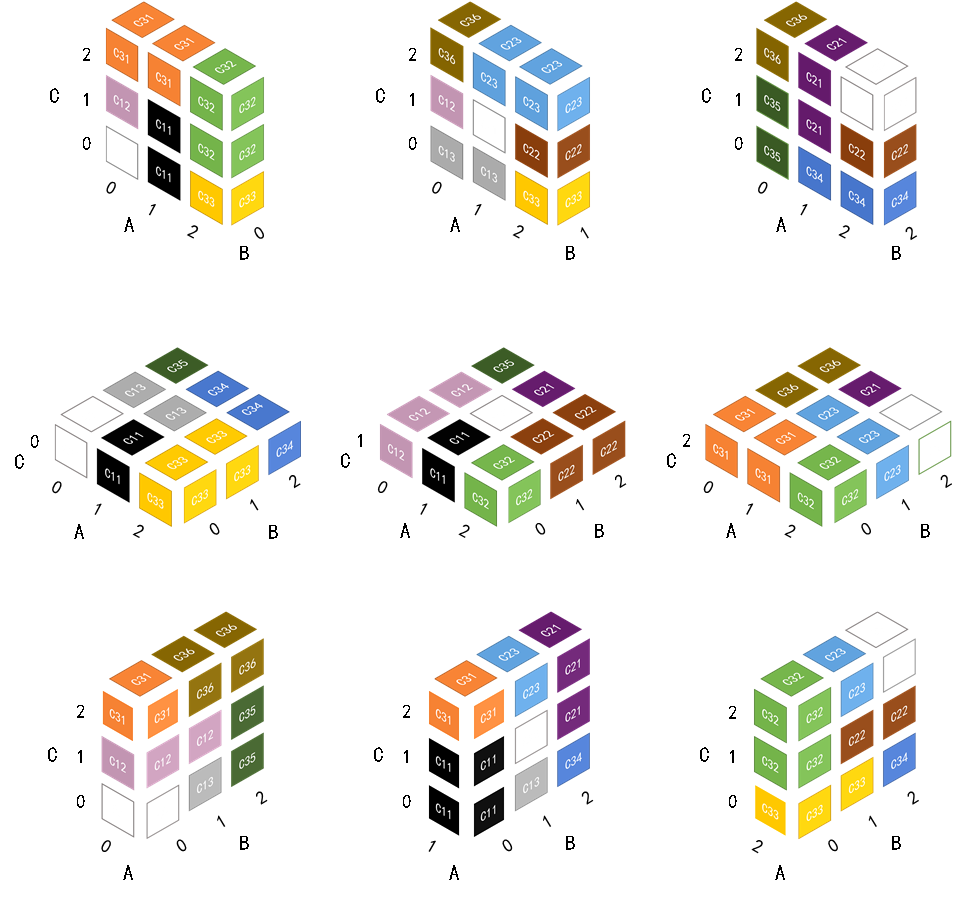}
	\caption{\label{fig:wide}all sections of the cube $3\times 3\times 3$.}
\end{figure*}

\textbf{Proposition 2: } If a 3-qubit UPB is strongly nonlocal, the tripartite state set satisfies the Tri-tile structure.

Proof: A 3-qubit UPB with strong nonlocality cannot locally eliminate one or more state while performing nontrivial orthogonality-preserving measurement. Hence, all tiles should be irreducible in ${{\mathcal {H} }_{A}}\otimes {{\mathcal {H} }_{B}}\otimes {{\mathcal {H} }_{C}}$ Hilbert space, which means any two tiles cannot be combined into a new tiles by simply shifting up and down or left and right. Therefore, this tripartite state set satisfies the Tri-tile structure.

According to the definition of Tri-tile structure, we construct an incomplete ${{C}^{3}}\times {{C}^{3}}\times {{C}^{4}}$ system based on Shifts UPB in Eq.(7), which is also strongly nonlocal. This system can lay the foundation for the further construction of 3-qubit UPB with higher dimension.

\begin{equation}
	\begin{aligned}
	  & {{C}_{1}}:\ \left\{ \begin{matrix}
		\left| {{\psi }_{0}} \right\rangle ={{\left| 1 \right\rangle }_{A}}{{\left| 0 \right\rangle }_{B}}{{\left| 0-1 \right\rangle }_{C}},  \\
		\left| {{\psi }_{1}} \right\rangle ={{\left| 0 \right\rangle }_{A}}{{\left| 0-1 \right\rangle }_{B}}{{\left| 1 \right\rangle }_{C}},  \\
		\left| {{\psi }_{2}} \right\rangle ={{\left| 0-1 \right\rangle }_{A}}{{\left| 1 \right\rangle }_{B}}{{\left| 0 \right\rangle }_{C}},  \\
	\end{matrix} \right. \\ 
	& {{C}_{2}}:\ \left\{ \begin{array}{*{35}{l}}
		\left| {{\psi }_{3}} \right\rangle ={{\left| 1 \right\rangle }_{A}}{{\left| 2 \right\rangle }_{B}}{{\left| 2-3 \right\rangle }_{C}},  \\
		\left| {{\psi }_{4}} \right\rangle ={{\left| 2 \right\rangle }_{A}}{{\left| 1-2 \right\rangle }_{B}}{{\left| 2 \right\rangle }_{C}},  \\
		\left| {{\psi }_{5}} \right\rangle ={{\left| 1-2 \right\rangle }_{A}}{{\left| 1 \right\rangle }_{B}}{{\left| 3 \right\rangle }_{C}},  \\
	\end{array} \right. \\ 
	& {{C}_{3}}:\ \left\{ \begin{matrix}
		\left| {{\psi }_{6}} \right\rangle ={{\left| 0-1 \right\rangle }_{A}}{{\left| 0-1 \right\rangle }_{B}}{{\left| 2 \right\rangle }_{C}}  \\
		\left| {{\psi }_{7}} \right\rangle ={{\left| 1-2 \right\rangle }_{A}}{{\left| 1-2 \right\rangle }_{B}}{{\left| 1 \right\rangle }_{C}}  \\
	\end{matrix} \right. \\ 
	& {{C}_{4}}:\ \left\{ \begin{array}{*{35}{l}}
		\left| {{\psi }_{8,9}} \right\rangle ={{\left| 2 \right\rangle }_{A}}{{\left| 0 \right\rangle }_{B}}{{\left| 1+w_{4}^{s}2+w_{4}^{2s}3 \right\rangle }_{C}}  \\
		\left| {{\psi }_{10}} \right\rangle ={{\left| 1-2 \right\rangle }_{A}}{{\left| 2 \right\rangle }_{B}}{{\left| 0 \right\rangle }_{C}}  \\
		\left| {{\psi }_{11}} \right\rangle ={{\left| 0 \right\rangle }_{A}}{{\left| 1-2 \right\rangle }_{B}}{{\left| 3 \right\rangle }_{C}}  \\
		\left| {{\psi }_{12,13}} \right\rangle ={{\left| 0 \right\rangle }_{A}}{{\left| 2 \right\rangle }_{B}}{{\left| 0+w_{4}^{s}1+w_{4}^{2s}2 \right\rangle }_{C}}  \\
		\left| {{\psi }_{14}} \right\rangle ={{\left| 0-1 \right\rangle }_{A}}{{\left| 0 \right\rangle }_{B}}{{\left| 3 \right\rangle }_{C}}  \\
		\left| {{\psi }_{15}} \right\rangle ={{\left| 2 \right\rangle }_{A}}{{\left| 0-1 \right\rangle }_{B}}{{\left| 0 \right\rangle }_{C}}  \\
	\end{array} \right. \\ 
	& Stopper:\ \left| {{\psi }_{16}} \right\rangle ={{\left( \left| 0 \right\rangle +\left| 1 \right\rangle +\left| 2 \right\rangle  \right)}_{A}}\otimes \\
	&{{\left( \left| 0 \right\rangle +\left| 1 \right\rangle +\left| 2 \right\rangle  \right)}_{B}}\otimes {{\left( \left| 0 \right\rangle +\left| 1 \right\rangle +\left| 2 \right\rangle +\left| 3 \right\rangle  \right)}_{C}}. \\ 
	\end{aligned}
\end{equation}

\subsection{Construct a UPB in ${{C}^{3}}\times {{C}^{3}}\times {{C}^{4}}$ system based on a UPB in ${{C}^{4}}\times {{C}^{4}}\times {{C}^{5}}$ system}
		
When constructing UPB of three parties with different dimensions, we adopt the similar practical method as before to construct high-dimensional state sets based on low-dimensional state sets. Though verifying whether the constructed state set meets the Tri-tile structure, we can judge whether the state set is strongly nonlocal more quickly and efficiently. Eq.(8) is a set of incomplete UPB in ${{C}^{4}}\times {{C}^{4}}\times {{C}^{5}}$.

\begin{equation}
	\begin{aligned}
	& {{C}_{1}}:\ \left\{ \begin{array}{*{35}{l}}
		\left| {{\psi }_{0}} \right\rangle ={{\left| 1 \right\rangle }_{A}}{{\left| 0 \right\rangle }_{B}}{{\left| 0-1 \right\rangle }_{C}},  \\
		\left| {{\psi }_{1}} \right\rangle ={{\left| 0 \right\rangle }_{A}}{{\left| 0-1 \right\rangle }_{B}}{{\left| 1 \right\rangle }_{C}},  \\
		\left| {{\psi }_{2}} \right\rangle ={{\left| 0-1 \right\rangle }_{A}}{{\left| 1 \right\rangle }_{B}}{{\left| 0 \right\rangle }_{C}},  \\
		\left| {{\psi }_{3}} \right\rangle ={{\left| 1 \right\rangle }_{A}}{{\left| 2 \right\rangle }_{B}}{{\left| 2-3 \right\rangle }_{C}},  \\
		\left| {{\psi }_{4}} \right\rangle ={{\left| 2 \right\rangle }_{A}}{{\left| 1-2 \right\rangle }_{B}}{{\left| 2 \right\rangle }_{C}},  \\
		\left| {{\psi }_{5}} \right\rangle ={{\left| 1-2 \right\rangle }_{A}}{{\left| 1 \right\rangle }_{B}}{{\left| 3 \right\rangle }_{C}},  \\
		\left| {{\psi }_{6,7}} \right\rangle ={{\left| 2 \right\rangle }_{A}}{{\left| 0 \right\rangle }_{B}}{{\left| 1+w_{4}^{s}2+w_{4}^{2s}3 \right\rangle }_{C}},  \\
		\left| {{\psi }_{8}} \right\rangle ={{\left| 1-2 \right\rangle }_{A}}{{\left| 2 \right\rangle }_{B}}{{\left| 0 \right\rangle }_{C}},  \\
		\left| {{\psi }_{9}} \right\rangle ={{\left| 0 \right\rangle }_{A}}{{\left| 1-2 \right\rangle }_{B}}{{\left| 3 \right\rangle }_{C}},  \\
		\left| {{\psi }_{10,11}} \right\rangle ={{\left| 0 \right\rangle }_{A}}{{\left| 2 \right\rangle }_{B}}{{\left| 0+w_{4}^{s}1+w_{4}^{2s}2 \right\rangle }_{C}},  \\
		\left| {{\psi }_{12}} \right\rangle ={{\left| 0-1 \right\rangle }_{A}}{{\left| 0 \right\rangle }_{B}}{{\left| 3 \right\rangle }_{C}},  \\
		\left| {{\psi }_{13}} \right\rangle ={{\left| 2 \right\rangle }_{A}}{{\left| 0-1 \right\rangle }_{B}}{{\left| 0 \right\rangle }_{C}},  \\
		\left| {{\psi }_{14}} \right\rangle ={{\left| 0-1 \right\rangle }_{A}}{{\left| 0-1 \right\rangle }_{B}}{{\left| 2 \right\rangle }_{C}},  \\
		\left| {{\psi }_{15}} \right\rangle ={{\left| 1-2 \right\rangle }_{A}}{{\left| 1-2 \right\rangle }_{B}}{{\left| 1 \right\rangle }_{C}},  \\
	\end{array} \right. \\ 
	& {{C}_{2}}:\ \left\{ \begin{array}{*{35}{l}}
		\left| {{\psi }_{16}} \right\rangle ={{\left| 3 \right\rangle }_{A}}{{\left| 2 \right\rangle }_{B}}{{\left| 3-4 \right\rangle }_{C}},  \\
		\left| {{\psi }_{17}} \right\rangle ={{\left| 2 \right\rangle }_{A}}{{\left| 2-3 \right\rangle }_{B}}{{\left| 4 \right\rangle }_{C}},  \\
		\left| {{\psi }_{18}} \right\rangle ={{\left| 2-3 \right\rangle }_{A}}{{\left| 3 \right\rangle }_{B}}{{\left| 3 \right\rangle }_{C}},  \\
	\end{array} \right. \\ 
	& {{C}_{3}}:\ \left\{ \begin{array}{*{35}{l}}
		\left| {{\psi }_{19}} \right\rangle ={{\left| 1 \right\rangle }_{A}}{{\left| 1-2 \right\rangle }_{B}}{{\left| 4 \right\rangle }_{C}},  \\
		\left| {{\psi }_{20}} \right\rangle ={{\left| 2-3 \right\rangle }_{A}}{{\left| 1 \right\rangle }_{B}}{{\left| 4 \right\rangle }_{C}},  \\
		\left| {{\psi }_{21}} \right\rangle ={{\left| 3 \right\rangle }_{A}}{{\left| 1 \right\rangle }_{B}}{{\left| 2-3 \right\rangle }_{C}},  \\
		\left| {{\psi }_{22,23}} \right\rangle ={{\left| 3 \right\rangle }_{A}}{{\left| 1+w_{4}^{s}2+w_{4}^{2s}3 \right\rangle }_{B}}{{\left| 1 \right\rangle }_{C}},  \\
		\left| {{\psi }_{24,25,26}} \right\rangle ={{\left| 1 \right\rangle }_{A}}{{\left| 3 \right\rangle }_{B}}{{\left| 1+w_{4}^{s}2+w_{4}^{2s}3+w_{4}^{3s}4 \right\rangle }_{C}},  \\
		\left| {{\psi }_{27}} \right\rangle ={{\left| 3 \right\rangle }_{A}}{{\left| 2-3 \right\rangle }_{B}}{{\left| 2 \right\rangle }_{C}},  \\
		\left| {{\psi }_{28}} \right\rangle ={{\left| 2 \right\rangle }_{A}}{{\left| 3 \right\rangle }_{B}}{{\left| 1-2 \right\rangle }_{C}},  \\
	\end{array} \right. \\ 
	& {{C}_{4}}:\ \left\{ \begin{array}{*{35}{l}}
		\left| {{\psi }_{29,30,31}} \right\rangle ={{\left| 0 \right\rangle }_{A}}{{\left| 3 \right\rangle }_{B}}{{\left| 0+w_{4}^{s}1+w_{4}^{2s}2+w_{4}^{3s}3 \right\rangle }_{C}},  \\
		\left| {{\psi }_{32,33}} \right\rangle ={{\left| 0+w_{4}^{s}1+w_{4}^{2s}2 \right\rangle }_{A}}{{\left| 0 \right\rangle }_{B}}{{\left| 4 \right\rangle }_{C}},  \\
		\left| {{\psi }_{34,35}} \right\rangle ={{\left| 3 \right\rangle }_{A}}{{\left| 0+w_{4}^{s}1+w_{4}^{2s}2 \right\rangle }_{B}}{{\left| 0 \right\rangle }_{C}},  \\
		\left| {{\psi }_{36,37,38}} \right\rangle ={{\left| 3 \right\rangle }_{A}}{{\left| 0 \right\rangle }_{B}}{{\left| 1+w_{4}^{s}2+w_{4}^{2s}3+w_{4}^{3s}4 \right\rangle }_{C}},  \\
		\left| {{\psi }_{39,40}} \right\rangle ={{\left| 1+w_{4}^{s}2+w_{4}^{2s}3 \right\rangle }_{A}}{{\left| 3 \right\rangle }_{B}}{{\left| 0 \right\rangle }_{C}},  \\
		\left| {{\psi }_{41,42}} \right\rangle ={{\left| 0 \right\rangle }_{A}}{{\left| 1+w_{4}^{s}2+w_{4}^{2s}3 \right\rangle }_{B}}{{\left| 4 \right\rangle }_{C}},  \\
	\end{array} \right. \\ 
	& Stopper:\ \left| {{\psi }_{43}} \right\rangle ={{\left( \left| 0 \right\rangle +\left| 1 \right\rangle +\left| 2 \right\rangle +\left| 3 \right\rangle  \right)}_{A}}\otimes  \\ 
	& {{\left( \left| 0 \right\rangle +\left| 1 \right\rangle +\left| 2 \right\rangle +\left| 3 \right\rangle  \right)}_{B}}\otimes {{\left( \left| 0 \right\rangle +\left| 1 \right\rangle +\left| 2 \right\rangle +\left| 3 \right\rangle  \right)}_{C}}. \\ 
\end{aligned}
\end{equation}

Similarly, the UPB of ${{C}^{4}}\times {{C}^{4}}\times {{C}^{5}}$ can be divided into the following four parts:

C1: The UPB system of ${{C}^{3}}\times {{C}^{3}}\times {{C}^{4}}$ in the lower left corner. It includes 14 subcubes and is nonlocal, as shown in Figure 11-a.

C2: The set of UPB with $d=2$in the upper right corner, which structure is the same as the Shifts UPB. It includes three subcubes and is nonlocal, as shown in Figure 11-b.

C3: The remaining parts of the three newly added planes. It can be combined with the $C_{3}^{d-1}$ and $C_{4}^{d-1}$ parts and decomposed into 7 subcubes that cannot be further expanded. This system is non-local, as shown in Figure 11-c.

C4: Six edges of the $4\times 4\times 5$ Rubik's Cube, corresponding to 6 subcubes. This system is nonlocal, as shown in Fig. 11-d.

\begin{figure*}
	\includegraphics[width=0.7\textwidth]{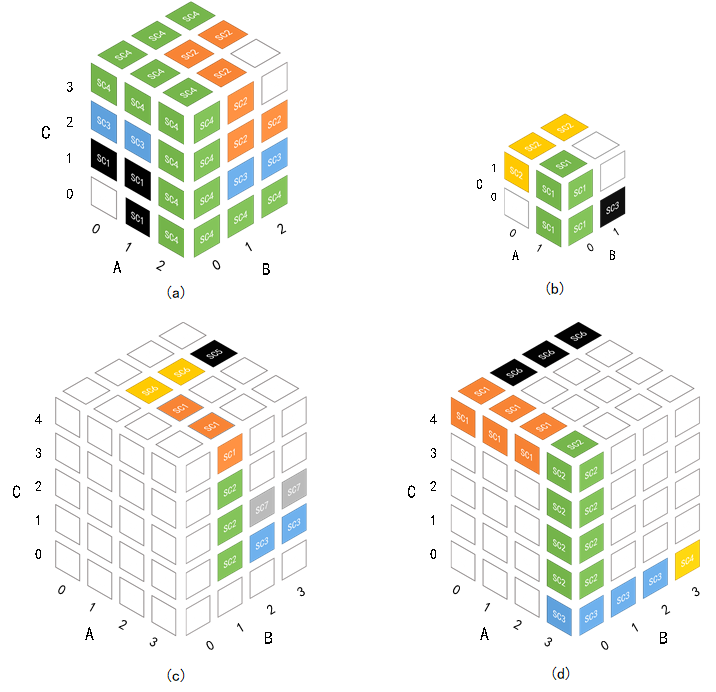}
	\caption{\label{fig:wide}composition of UPB in ${{C}^{4}}\otimes {{C}^{4}}\otimes {{C}^{5}}$.}
\end{figure*}

\subsection{Construct a UPB in ${{C}^{{{d}_{1}}}}\otimes {{C}^{{{d}_{2}}}}\otimes {{C}^{{{d}_{3}}}}$ system based on a UPB in ${{C}^{\left( {{d}_{1}}-1 \right)}}\otimes {{C}^{\left( {{d}_{2}}-1 \right)}}\otimes {{C}^{\left( {{d}_{3}}-1 \right)}}$ system}

In this section, preserving the tile structure of the UPB in ${{C}^{\left( {{d}_{1}}-1 \right)}}\otimes {{C}^{\left( {{d}_{2}}-1 \right)}}\otimes {{C}^{\left( {{d}_{3}}-1 \right)}}$,  we propose a general construction method of UPB in ${{C}^{{{d}_{1}}}}\otimes {{C}^{{{d}_{2}}}}\otimes {{C}^{{{d}_{3}}}},\quad d{}_{1},\ d{}_{2},\ d{}_{3}\,\in \left[ 0,\ d-1 \right]$ according to the proposed Tri-tile structure.

Continuing with the aforesaid construction method, we first divide the six edges into a group of six states that cannot be further expanded. In addition to the set in ${{C}^{\left( {{d}_{1}}-1 \right)}}\otimes {{C}^{\left( {{d}_{2}}-1 \right)}}\otimes {{C}^{\left( {{d}_{3}}-1 \right)}}$ system and a set of Shift UPB, there will be some bases left in the outermost layers. Then, for the remaining bases, it is necessary to ensure that the decomposition method of newly added states can satisfies the Tri-tile structure. 

\textbf{Proposition 3: }Let ${{\mathcal {H} }_{A}},\ {{\mathcal {H} }_{B}},\ {{\mathcal {H} }_{C}}$ be Hilbert spaces of dimension x, y, z respectively. Suppose that $A=\left( {{\left| 1 \right\rangle }_{A}},\cdots ,\ \ {{\left| {{d}_{1}} \right\rangle }_{A}} \right)$,  $B=\left( {{\left| 1 \right\rangle }_{B}},\cdots ,\ \ {{\left| {{d}_{2}} \right\rangle }_{B}} \right)$,  $C=\left( {{\left| 1 \right\rangle }_{C}},\cdots ,\ \ {{\left| {{d}_{3}} \right\rangle }_{C}} \right)$ are ordered orthonormal bases with respect to ${{\mathcal {H} }_{A}},\ {{\mathcal {H} }_{B}},\ {{\mathcal {H} }_{C}}$. In \[{{C}^{{{d}_{1}}}}\otimes {{C}^{{{d}_{2}}}}\otimes {{C}^{{{d}_{3}}}}\], the set of UPB given by Eq. (9) is pairwise orthogonal and strongly nonlocal.
\begin{widetext}
\begin{equation}
\begin{aligned}
		& {{C}_{1}}:\ C_{1}^{d-1}\otimes C_{1}^{d-1}\otimes C_{1}^{d-1} \\ 
		& {{C}_{2}}:\ C_{2}^{d-1}+1 \\ 
		& {{C}_{4}}:\ \left\{ \begin{array}{*{35}{l}}
			\left| {{\psi }_{k}} \right\rangle ={{\left| 0+\cdots +w_{d}^{\left( d-2 \right)s}\left( d-2 \right) \right\rangle }_{A}}{{\left| 0 \right\rangle }_{B}}{{\left| i \right\rangle }_{C}}  \\
			\left| {{\psi }_{k+d-2}} \right\rangle ={{\left| i \right\rangle }_{A}}{{\left| 0+\cdots +w_{d}^{\left( d-2 \right)s}\left( d-2 \right) \right\rangle }_{B}}{{\left| 0 \right\rangle }_{C}}  \\
			\left| {{\psi }_{k+2\left( d-2 \right)}} \right\rangle ={{\left| 0 \right\rangle }_{A}}{{\left| i \right\rangle }_{B}}{{\left| 0+\cdots +w_{d}^{\left( d-2 \right)s}\left( d-2 \right) \right\rangle }_{C}}  \\
			\left| {{\psi }_{k+3\left( d-2 \right)}} \right\rangle ={{\left| 1+\cdots +w_{d}^{\left( d-1 \right)s}\left( d-1 \right) \right\rangle }_{A}}{{\left| i \right\rangle }_{B}}{{\left| 0 \right\rangle }_{C}}  \\
			\left| {{\psi }_{k+3\left( d-2 \right)+\left( d-1 \right)}} \right\rangle ={{\left| 0 \right\rangle }_{A}}{{\left| 1+\cdots +w_{d}^{\left( d-1 \right)s}\left( d-1 \right) \right\rangle }_{B}}{{\left| i \right\rangle }_{C}}  \\
			\left| {{\psi }_{k+3\left( d-2 \right)+2\left( d-1 \right)}} \right\rangle ={{\left| i \right\rangle }_{A}}{{\left| 0 \right\rangle }_{B}}{{\left| 1+\cdots +w_{d}^{\left( d-1 \right)s}\left( d-1 \right) \right\rangle }_{C}}  \\
		\end{array} \right. \\ 
		& Stopper:\ \left| {{\psi }_{k+3\left( d-2 \right)+3\left( d-1 \right)}} \right\rangle ={{\left( \left| 0 \right\rangle +\cdots +\left| d-1 \right\rangle  \right)}_{A}}{{\left( \left| 0 \right\rangle +\cdots +\left| d-1 \right\rangle  \right)}_{B}}{{\left( \left| 0 \right\rangle +\cdots +\left| d-1 \right\rangle  \right)}_{C}}. \\ 
\end{aligned}
\end{equation}
\end{widetext}

By mapping the set Eq. (9) to the Rubik's Cube, the structure is shown in Fig. 12.

\begin{figure*}
	\includegraphics[width=0.45\textwidth]{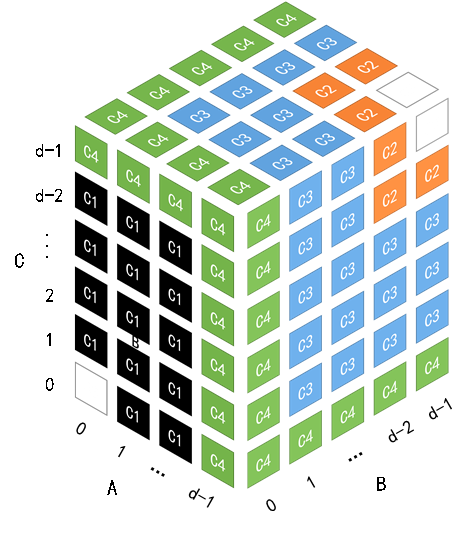}
	\caption{\label{fig:wide} composition of UPB in ${{C}^{{{d}_{1}}}}\otimes {{C}^{{{d}_{2}}}}\otimes {{C}^{{{d}_{3}}}}$.}
\end{figure*}

The UPB of ${{C}^{{{d}_{1}}}}\otimes {{C}^{{{d}_{2}}}}\otimes {{C}^{{{d}_{3}}}}$ can be divided into the following four parts:

C1: The UPB system of ${{C}^{\left( {{d}_{1}}-1 \right)}}\otimes {{C}^{\left( {{d}_{2}}-1 \right)}}\otimes {{C}^{\left( {{d}_{3}}-1 \right)}}$ in the lower left corner. 

C2: The set of UPB with $d=2$ in the upper right corner, which structure is the same as the Shifts UPB. It includes three subcubes and is nonlocal.

C3: The remaining parts of the three newly added planes. This part is hard to show in a definite expression. When constructing this part, It needs to ensure that the structure of state set meets the Tri-tile structure after this remaining part combined with the $C_{3}^{d-1}$ and $C_{4}^{d-1}$ parts. 

C4: Six edges of the ${{d}_{1}}\otimes {{d}_{2}}\otimes {{d}_{3}}$ Rubik's Cube, corresponding to 6 subcubes.

\section{\label{sec:level2}Conclusion}

In summary, we concentrated on the general construction method of 3-qubit UPB exhibiting strong nonlocality. Firstly, a strongly nonlocal set ${{C}^{3}}\otimes {{C}^{3}}\otimes {{C}^{3}}$ of size 12 was presented based on the Shifts UPB. After a simple observation of its orthogonal graphs and corresponding $3\times 3\times 3$ Rubik's cube, the relationship between three qubits was discussed and the structure of UPB was analyzed. Following the idea of deducing the high-dimensional UPB from the low-dimensional UPB, we tried to construct the set with higher dimensions and obtained a general set ${{C}^{d}}\otimes {{C}^{d}}\otimes {{C}^{d}}$ of size ${{\left( d-1 \right)}^{3}}+3\left( d-2 \right)+1$. Second, we extended the 2-qubit tile structure to 3-qubit and defined it as a Tri-tile structure, which is conducive for us to judge whether the state set is strongly nonlocal more quickly and efficiently. If the state set satisfies this structure, each section in the corresponding Rubik's Cube conforms to the tile structure. Then on this basis, we gave the construction process of 3-qubit UPB in ${{C}^{{{d}_{1}}}}\otimes {{C}^{{{d}_{2}}}}\otimes {{C}^{{{d}_{3}}}}$ system. The structure of it is similar to that of ${{C}^{d}}\otimes {{C}^{d}}\otimes {{C}^{d}}$ system, which also consists of four parts.

It is noted that the state sets which are constructed using the proposed general construction method include a considerable number of states. Considering that reducing the number of quantum states is of great significance for exploring which states affect the nonlocality of the system, it is a valuable research direction to discuss the minimum number of states in 3-qubit UPB. In addition, there is another kind of state that is closely related to UPB, bound entangled states (or BE states). Since UPBs proposed in this paper have a large size, they can produce BE states of small rank. The characteristics of BE states is also a worthy research direction.

\begin{acknowledgments}
This work is supported by the National Key R\&D Program of China (Grant Nos.  2020YFB1805405), the Foundation of Guizhou Provincial Key Laboratory of Public Big Data (Grant No. 2019BDKFJJ014) and the Fundamental Research Funds for the Central Universities (Grant Nos. 2019XD-A02, 2020RC38).
\end{acknowledgments}

\appendix

\section{Proof of Lemma 1}
Lemma 2: The following 34 states Eq. (5) is strongly nonlocal.

Proof: In order to make the proof process clearer, we first rewrite the original state, let $\left| 00 \right\rangle \to \left| 0 \right\rangle ,\ \left| 01 \right\rangle \to \left| 1 \right\rangle ,\ \cdots ,\ \left| 33 \right\rangle \to \left| 15 \right\rangle $. After rewriting the state Eq. (5), we can get the following states Eq. (A1):

\begin{equation}
	\begin{aligned}
		& {{C}_{1}}:\ \left\{ \begin{matrix}
			\left| {{\psi }_{0}} \right\rangle ={{\left| 1 \right\rangle }_{A}}{{\left| 0-1 \right\rangle }_{BC}},  \\
			\left| {{\psi }_{1}} \right\rangle ={{\left| 0 \right\rangle }_{A}}{{\left| 1-5 \right\rangle }_{BC}},  \\
			\left| {{\psi }_{2}} \right\rangle ={{\left| 0-1 \right\rangle }_{A}}{{\left| 4 \right\rangle }_{BC}},  \\
			\left| {{\psi }_{3}} \right\rangle ={{\left| 1 \right\rangle }_{A}}{{\left| 9-10 \right\rangle }_{BC}},  \\
			\left| {{\psi }_{4}} \right\rangle ={{\left| 2 \right\rangle }_{A}}{{\left| 5-9 \right\rangle }_{BC}},  \\
			\left| {{\psi }_{5}} \right\rangle ={{\left| 1-2 \right\rangle }_{A}}{{\left| 6 \right\rangle }_{BC}},  \\
			\left| {{\psi }_{6}} \right\rangle ={{\left| 2 \right\rangle }_{A}}{{\left| 1-2 \right\rangle }_{BC}},  \\
			\left| {{\psi }_{7}} \right\rangle ={{\left| 1-2 \right\rangle }_{A}}{{\left| 8 \right\rangle }_{BC}},  \\
			\left| {{\psi }_{8}} \right\rangle ={{\left| 0 \right\rangle }_{A}}{{\left| 6-10 \right\rangle }_{BC}},  \\
			\left| {{\psi }_{9}} \right\rangle ={{\left| 0 \right\rangle }_{A}}{{\left| 8-9 \right\rangle }_{BC}},  \\
			\left| {{\psi }_{10}} \right\rangle ={{\left| 0-1 \right\rangle }_{A}}{{\left| 2 \right\rangle }_{BC}},  \\
			\left| {{\psi }_{11}} \right\rangle ={{\left| 2 \right\rangle }_{A}}{{\left| 0-4 \right\rangle }_{BC}},  \\
		\end{matrix} \right. \\ 
		& {{C}_{2}}:\ \left\{ \begin{matrix}
			\left| {{\psi }_{12}} \right\rangle ={{\left| 3 \right\rangle }_{A}}{{\left| 10-11 \right\rangle }_{BC}},  \\
			\left| {{\psi }_{13}} \right\rangle ={{\left| 2 \right\rangle }_{A}}{{\left| 11-15 \right\rangle }_{BC}},  \\
			\left| {{\psi }_{14}} \right\rangle ={{\left| 2-3 \right\rangle }_{A}}{{\left| 14 \right\rangle }_{BC}},  \\
		\end{matrix} \right. \\ 
		& {{C}_{3}}:\ \left\{ \begin{matrix}
			\left| {{\psi }_{15}} \right\rangle ={{\left| 1 \right\rangle }_{A}}{{\left| 14-15 \right\rangle }_{BC}},  \\
			\left| {{\psi }_{16}} \right\rangle ={{\left| 3 \right\rangle }_{A}}{{\left| 9-13 \right\rangle }_{BC}},  \\
			\left| {{\psi }_{17}} \right\rangle ={{\left| 2-3 \right\rangle }_{A}}{{\left| 7 \right\rangle }_{BC}},  \\
			\left| {{\psi }_{18}} \right\rangle ={{\left| 3 \right\rangle }_{A}}{{\left| 5-6 \right\rangle }_{BC}},  \\
			\left| {{\psi }_{19}} \right\rangle ={{\left| 1 \right\rangle }_{A}}{{\left| 7-11 \right\rangle }_{BC}},  \\
			\left| {{\psi }_{20}} \right\rangle ={{\left| 1-2 \right\rangle }_{A}}{{\left| 13 \right\rangle }_{BC}},  \\
		\end{matrix} \right. \\ 
		& {{C}_{4}}:\ \left\{ \begin{array}{*{35}{l}}
			\left| {{\psi }_{21,22}} \right\rangle ={{\left| 0 \right\rangle }_{A}}{{\left| 12+w_{4}^{s}13+w_{4}^{2s}14 \right\rangle }_{BC}},  \\
			\left| {{\psi }_{23.24}} \right\rangle ={{\left| 0+w_{4}^{s}1+w_{4}^{2s}2 \right\rangle }_{A}}{{\left| 3 \right\rangle }_{BC}},  \\
			\left| {{\psi }_{25,26}} \right\rangle ={{\left| 3 \right\rangle }_{A}}{{\left| 0+w_{4}^{s}4+w_{4}^{2s}8 \right\rangle }_{BC}},  \\
			\left| {{\psi }_{27,28}} \right\rangle ={{\left| 3 \right\rangle }_{A}}{{\left| 1+w_{4}^{s}2+w_{4}^{2s}3 \right\rangle }_{BC}},  \\
			\left| {{\psi }_{29.30}} \right\rangle ={{\left| 1+w_{4}^{s}2+w_{4}^{2s}3 \right\rangle }_{A}}{{\left| 12 \right\rangle }_{BC}},  \\
			\left| {{\psi }_{31,32}} \right\rangle ={{\left| 0 \right\rangle }_{A}}{{\left| 7+w_{4}^{s}11+w_{4}^{2s}15 \right\rangle }_{BC}},  \\
		\end{array} \right. \\ 
		& Stopper:\ \left| {{\psi }_{33}} \right\rangle ={{\left( \left| 0 \right\rangle +\left| 1 \right\rangle +\left| 2 \right\rangle +\left| 3 \right\rangle  \right)}_{A}}{{\left( \left| 0 \right\rangle +\cdots +\left| 15 \right\rangle  \right)}_{B}}_{C}. \\ 
	\end{aligned}
\end{equation}

Suppose \textit{BC} system starts with the nontrivial and non-disturbing measurement, represented by a set of POVM elements $M_{m}^{\dagger }M_{m}^{{}}$ on ${{d}^{2}}\times {{d}^{2}}$. The POVM measurement in ${{\left\{ \left| 0 \right\rangle ,\left| 1 \right\rangle ,\cdots ,\left| 15 \right\rangle  \right\}}_{A}}$ basis can be written, which corresponds to the states Eq. (A1):

\begin{eqnarray*}
	M_{m}^{\dagger }M_{m}^{{}}=\left[ \begin{matrix}
		{{a}_{00}} & {{a}_{01}} & \cdots  & {{a}_{0\left( 14 \right)}} & {{a}_{0\left( 15 \right)}}  \\
		{{a}_{10}} & {{a}_{11}} & \cdots  & {{a}_{1\left( 14 \right)}} & {{a}_{1\left( 15 \right)}}  \\
		\vdots  & \vdots  & \ddots  & \vdots  & \vdots   \\
		{{a}_{\left( 14 \right)0}} & {{a}_{\left( 14 \right)1}} & \cdots  & {{a}_{\left( 14 \right)\left( 14 \right)}} & {{a}_{\left( 14 \right)\left( 15 \right)}}  \\
		{{a}_{\left( 15 \right)0}} & {{a}_{\left( 15 \right)1}} & \cdots  & {{a}_{\left( 15 \right)\left( 14 \right)}} & {{a}_{\left( 15 \right)\left( 15 \right)}}  \\
	\end{matrix} \right]
\end{eqnarray*}

The post-measurement states could be expressed as $\left( I\otimes {{M}_{m}} \right)\left| {{\varphi }_{i}} \right\rangle $, which should be mutually orthogonal. Then $\left\langle  {{\varphi }_{j}} \right|\left( I\otimes M_{m}^{\dagger }M_{m}^{{}} \right)\left| {{\varphi }_{i}} \right\rangle =0$ is obtained. According to this principle, the original matrix could be transformed into:

\begin{eqnarray*}
	M_{m}^{\dagger }M_{m}^{{}}=\left[ \begin{matrix}
		a & 0 & \cdots  & 0  \\
		0 & a & \cdots  & 0  \\
		\vdots  & \vdots  & \ddots  & \vdots   \\
		0 & 0 & \cdots  & {{a}_{{}}}  \\
	\end{matrix} \right]
\end{eqnarray*}

Table II shows the detailed derivation process.
\begin{table*}
	\caption{\label{tab:tableA1} POVM elements}
	\begin{ruledtabular}
		\begin{tabular}{cccccc}
			POVM Element & \multicolumn{5}{c}{Corresponding States} \\
			\hline
			\multirow{6}{*}{${{a}_{0i}}={{a}_{i0}}=0$} & $i=1$ & $i=2$ & $i=3$ & $i=4$ & $i=5$\\
			&$\left| {{\varphi }_{11}} \right\rangle $,$\left| {{\varphi }_{1}} \right\rangle $ &$\left| {{\varphi }_{11}} \right\rangle $,$\left| {{\varphi }_{10}} \right\rangle $ &$\left| {{\varphi }_{11}} \right\rangle $,$\left| {{\varphi }_{22}} \right\rangle $ &$\left| {{\varphi }_{0}} \right\rangle $,$\left| {{\varphi }_{2}} \right\rangle $ &$\left| {{\varphi }_{11}} \right\rangle $,$\left| {{\varphi }_{4}} \right\rangle $ \\
			&$i=6$ &$i=7$ &$i=8$ &$i=9$ &$i=10$\\
			&$\left| {{\varphi }_{11}} \right\rangle $,$\left| {{\varphi }_{5}} \right\rangle $ &$\left| {{\varphi }_{11}} \right\rangle $,$\left| {{\varphi }_{17}} \right\rangle $ &$\left| {{\varphi }_{11}} \right\rangle $,$\left| {{\varphi }_{7}} \right\rangle $ &$\left| {{\varphi }_{11}} \right\rangle $,$\left| {{\varphi }_{9}} \right\rangle $ &$\left| {{\varphi }_{11}} \right\rangle $,$\left| {{\varphi }_{8}} \right\rangle $\\
			&$i=11$ &$i=12$ &$i=13$ &$i=14$ &$i=15$ \\
			&$\left| {{\varphi }_{11}} \right\rangle $,$\left| {{\varphi }_{12}} \right\rangle $ &$\left| {{\varphi }_{11}} \right\rangle $,$\left| {{\varphi }_{30}} \right\rangle $ &$\left| {{\varphi }_{11}} \right\rangle $,$\left| {{\varphi }_{20}} \right\rangle $ &$\left| {{\varphi }_{11}} \right\rangle $,$\left| {{\varphi }_{14}} \right\rangle $ &$\left| {{\varphi }_{11}} \right\rangle $,$\left| {{\varphi }_{13}} \right\rangle $ \\
			\hline
			\multirow{6}{*}{${{a}_{1i}}={{a}_{i1}}=0$} & $i=2$ & $i=3$ & $i=4$ & $i=5$ &$i=6$\\
			&$\left| {{\varphi }_{1}} \right\rangle $,$\left| {{\varphi }_{10}} \right\rangle $ &$\left| {{\varphi }_{0}} \right\rangle $,$\left| {{\varphi }_{24}} \right\rangle $ &$\left| {{\varphi }_{0}} \right\rangle $,$\left| {{\varphi }_{2}} \right\rangle $ &$\left| {{\varphi }_{0}} \right\rangle $,$\left| {{\varphi }_{18}} \right\rangle $ &$\left| {{\varphi }_{0}} \right\rangle $,$\left| {{\varphi }_{5}} \right\rangle $\\
			&$i=7$ &$i=8$ &$i=9$ &$i=10$ &$i=11$ \\
	    	&$\left| {{\varphi }_{0}} \right\rangle $,$\left| {{\varphi }_{17}} \right\rangle $ &$\left| {{\varphi }_{0}} \right\rangle $,$\left| {{\varphi }_{7}} \right\rangle $ &$\left| {{\varphi }_{0}} \right\rangle $,$\left| {{\varphi }_{9}} \right\rangle $ &$\left| {{\varphi }_{0}} \right\rangle $,$\left| {{\varphi }_{8}} \right\rangle $ &$\left| {{\varphi }_{0}} \right\rangle $,$\left| {{\varphi }_{12}} \right\rangle $ \\
			&$i=12$ &$i=13$ &$i=14$ &$i=15$ & \\
			&$\left| {{\varphi }_{0}} \right\rangle $,$\left| {{\varphi }_{30}} \right\rangle $ &$\left| {{\varphi }_{0}} \right\rangle $,$\left| {{\varphi }_{20}} \right\rangle $ &$\left| {{\varphi }_{0}} \right\rangle $,$\left| {{\varphi }_{14}} \right\rangle $ &$\left| {{\varphi }_{0}} \right\rangle $,$\left| {{\varphi }_{13}} \right\rangle $ & \\
			\hline
			\multirow{6}{*}{${{a}_{2i}}={{a}_{i2}}=0$} & $i=3$ & $i=4$ & $i=5$ & $i=6$ & $i=7$\\
			&$\left| {{\varphi }_{10}} \right\rangle $,$\left| {{\varphi }_{24}} \right\rangle $ &$\left| {{\varphi }_{10}} \right\rangle $,$\left| {{\varphi }_{2}} \right\rangle $ &$\left| {{\varphi }_{10}} \right\rangle $,$\left| {{\varphi }_{4}} \right\rangle $ &$\left| {{\varphi }_{10}} \right\rangle $,$\left| {{\varphi }_{5}} \right\rangle $ &$\left| {{\varphi }_{10}} \right\rangle $,$\left| {{\varphi }_{17}} \right\rangle $ \\
			& $i=8$ &$i=9$ &$i=10$ &$i=11$ &$i=12$ \\
			&$\left| {{\varphi }_{10}} \right\rangle $,$\left| {{\varphi }_{7}} \right\rangle $ &$\left| {{\varphi }_{10}} \right\rangle $,$\left| {{\varphi }_{9}} \right\rangle $ &$\left| {{\varphi }_{10}} \right\rangle $,$\left| {{\varphi }_{8}} \right\rangle $&$\left| {{\varphi }_{10}} \right\rangle $,$\left| {{\varphi }_{12}} \right\rangle $ &$\left| {{\varphi }_{10}} \right\rangle $,$\left| {{\varphi }_{30}} \right\rangle $\\
			&$i=13$ &$i=14$ &$i=15$ & & \\
			&$\left| {{\varphi }_{10}} \right\rangle $,$\left| {{\varphi }_{20}} \right\rangle $ &$\left| {{\varphi }_{10}} \right\rangle $,$\left| {{\varphi }_{14}} \right\rangle $ &$\left| {{\varphi }_{10}} \right\rangle $,$\left| {{\varphi }_{13}} \right\rangle $ & & \\
			\hline
			\multirow{6}{*}{${{a}_{3i}}={{a}_{i3}}=0$} & $i=4$ & $i=5$ & $i=6$ & $i=7$& $i=8$ \\
			&$\left| {{\varphi }_{24}} \right\rangle $,$\left| {{\varphi }_{2}} \right\rangle $ &$\left| {{\varphi }_{24}} \right\rangle $,$\left| {{\varphi }_{1}} \right\rangle $ 
			&$\left| {{\varphi }_{24}} \right\rangle $,$\left| {{\varphi }_{5}} \right\rangle $ &$\left| {{\varphi }_{24}} \right\rangle $,$\left| {{\varphi }_{17}} \right\rangle $ &$\left| {{\varphi }_{24}} \right\rangle $,$\left| {{\varphi }_{7}} \right\rangle $ \\
			&$i=9$ &$i=10$ &$i=11$	&$i=12$ &$i=13$ \\
			&$\left| {{\varphi }_{24}} \right\rangle $,$\left| {{\varphi }_{9}} \right\rangle $ &$\left| {{\varphi }_{24}} \right\rangle $,$\left| {{\varphi }_{8}} \right\rangle $ &$\left| {{\varphi }_{24}} \right\rangle $,$\left| {{\varphi }_{12}} \right\rangle $&$\left| {{\varphi }_{24}} \right\rangle $,$\left| {{\varphi }_{30}} \right\rangle $ &$\left| {{\varphi }_{24}} \right\rangle $,$\left| {{\varphi }_{20}} \right\rangle $ \\
			&$i=14$ &$i=15$ & & & \\
			&$\left| {{\varphi }_{24}} \right\rangle $,$\left| {{\varphi }_{14}} \right\rangle $ &$\left| {{\varphi }_{24}} \right\rangle $,$\left| {{\varphi }_{13}} \right\rangle $ & & & \\
			\hline
			\multirow{6}{*}{${{a}_{4i}}={{a}_{i4}}=0$} & $i=5$ & $i=6$ & $i=7$& $i=8$ &$i=9$ \\
			&$\left| {{\varphi }_{2}} \right\rangle $,$\left| {{\varphi }_{18}} \right\rangle $ &$\left| {{\varphi }_{2}} \right\rangle $,$\left| {{\varphi }_{5}} \right\rangle $ 
			&$\left| {{\varphi }_{2}} \right\rangle $,$\left| {{\varphi }_{17}} \right\rangle $ &$\left| {{\varphi }_{2}} \right\rangle $,$\left| {{\varphi }_{7}} \right\rangle $ &$\left| {{\varphi }_{2}} \right\rangle $,$\left| {{\varphi }_{4}} \right\rangle $ \\
			&$i=10$ &$i=11$ &$i=12$ &$i=13$ &$i=14$ \\
			&$\left| {{\varphi }_{2}} \right\rangle $,$\left| {{\varphi }_{8}} \right\rangle $ &$\left| {{\varphi }_{2}} \right\rangle $,$\left| {{\varphi }_{12}} \right\rangle $ &$\left| {{\varphi }_{2}} \right\rangle $,$\left| {{\varphi }_{30}} \right\rangle $&$\left| {{\varphi }_{2}} \right\rangle $,$\left| {{\varphi }_{20}} \right\rangle $ &$\left| {{\varphi }_{2}} \right\rangle $,$\left| {{\varphi }_{14}} \right\rangle $\\
			&$i=15$ & & & & \\
			&$\left| {{\varphi }_{2}} \right\rangle $,$\left| {{\varphi }_{13}} \right\rangle $ & & & & \\
			\hline
			\multirow{4}{*}{${{a}_{5i}}={{a}_{i5}}=0$} & $i=6$ & $i=7$& $i=8$ &$i=9$ &$i=10$ \\
			&$\left| {{\varphi }_{1}} \right\rangle $,$\left| {{\varphi }_{5}} \right\rangle $ &$\left| {{\varphi }_{18}} \right\rangle $,$\left| {{\varphi }_{17}} \right\rangle $ &$\left| {{\varphi }_{1}} \right\rangle $,$\left| {{\varphi }_{7}} \right\rangle $ &$\left| {{\varphi }_{1}} \right\rangle $,$\left| {{\varphi }_{9}} \right\rangle $ &$\left| {{\varphi }_{1}} \right\rangle $,$\left| {{\varphi }_{8}} \right\rangle $ \\
			&$i=11$ &$i=12$ &$i=13$ &$i=14$ &$i=15$ \\
			&$\left| {{\varphi }_{1}} \right\rangle $,$\left| {{\varphi }_{12}} \right\rangle $ &$\left| {{\varphi }_{1}} \right\rangle $,$\left| {{\varphi }_{30}} \right\rangle $ &$\left| {{\varphi }_{1}} \right\rangle $,$\left| {{\varphi }_{20}} \right\rangle $ &$\left| {{\varphi }_{1}} \right\rangle $,$\left| {{\varphi }_{14}} \right\rangle $ &$\left| {{\varphi }_{1}} \right\rangle $,$\left| {{\varphi }_{13}} \right\rangle $ \\
			\hline
			\multirow{4}{*}{${{a}_{6i}}={{a}_{i6}}=0$} & $i=7$& $i=8$ &$i=9$ &$i=10$ &$i=11$ \\
			&$\left| {{\varphi }_{5}} \right\rangle $,$\left| {{\varphi }_{17}} \right\rangle $ &$\left| {{\varphi }_{5}} \right\rangle $,$\left| {{\varphi }_{7}} \right\rangle $ &$\left| {{\varphi }_{5}} \right\rangle $,$\left| {{\varphi }_{4}} \right\rangle $ &$\left| {{\varphi }_{5}} \right\rangle $,$\left| {{\varphi }_{3}} \right\rangle $ &$\left| {{\varphi }_{5}} \right\rangle $,$\left| {{\varphi }_{12}} \right\rangle $\\
			&$i=12$ &$i=13$ &$i=14$	&$i=15$ & \\
			&$\left| {{\varphi }_{5}} \right\rangle $,$\left| {{\varphi }_{30}} \right\rangle $ &$\left| {{\varphi }_{5}} \right\rangle $,$\left| {{\varphi }_{20}} \right\rangle $ &$\left| {{\varphi }_{5}} \right\rangle $,$\left| {{\varphi }_{14}} \right\rangle $ &$\left| {{\varphi }_{5}} \right\rangle $,$\left| {{\varphi }_{13}} \right\rangle $ &\\
			\hline
			\multirow{4}{*}{${{a}_{7i}}={{a}_{i7}}=0$} &$i=8$ &$i=9$ &$i=10$ &$i=11$ &$i=12$ \\
			&$\left| {{\varphi }_{17}} \right\rangle $,$\left| {{\varphi }_{7}} \right\rangle$ &$\left| {{\varphi }_{17}} \right\rangle $,$\left| {{\varphi }_{9}} \right\rangle $ &$\left| {{\varphi }_{17}} \right\rangle $,$\left| {{\varphi }_{8}} \right\rangle $ &$\left| {{\varphi }_{17}} \right\rangle $,$\left| {{\varphi }_{12}} \right\rangle $ &$\left| {{\varphi }_{17}} \right\rangle $,$\left| {{\varphi }_{30}} \right\rangle $ \\
			&$i=13$ &$i=14$ &$i=15$ & &\\
			&$\left| {{\varphi }_{17}} \right\rangle $,$\left| {{\varphi }_{20}} \right\rangle $ &$\left| {{\varphi }_{17}} \right\rangle $,$\left| {{\varphi }_{14}} \right\rangle $ &$\left| {{\varphi }_{17}} \right\rangle $,$\left| {{\varphi }_{13}} \right\rangle $ & &\\
			\hline
			\multirow{4}{*}{${{a}_{8i}}={{a}_{i8}}=0$} &$i=9$ &$i=10$ &$i=11$ &$i=12$ &$i=13$ \\
			&$\left| {{\varphi }_{7}} \right\rangle $,$\left| {{\varphi }_{4}} \right\rangle$ &$\left| {{\varphi }_{7}} \right\rangle $,$\left| {{\varphi }_{8}} \right\rangle $ &$\left| {{\varphi }_{7}} \right\rangle $,$\left| {{\varphi }_{12}} \right\rangle $ &$\left| {{\varphi }_{7}} \right\rangle $,$\left| {{\varphi }_{30}} \right\rangle $ &$\left| {{\varphi }_{7}} \right\rangle $,$\left| {{\varphi }_{20}} \right\rangle $ \\
			&$i=14$ &$i=15$ & & &\\
			&$\left| {{\varphi }_{7}} \right\rangle $,$\left| {{\varphi }_{14}} \right\rangle $ &$\left| {{\varphi }_{7}} \right\rangle $,$\left| {{\varphi }_{13}} \right\rangle $ & & & \\
			\hline
			\multirow{4}{*}{${{a}_{9i}}={{a}_{i9}}=0$} &$i=10$ &$i=11$ &$i=12$ &$i=13$ &$i=14$ \\
			&$\left| {{\varphi }_{4}} \right\rangle $,$\left| {{\varphi }_{8}} \right\rangle $ &$\left| {{\varphi }_{4}} \right\rangle $,$\left| {{\varphi }_{12}} \right\rangle $ &$\left| {{\varphi }_{4}} \right\rangle $,$\left| {{\varphi }_{30}} \right\rangle $ &$\left| {{\varphi }_{4}} \right\rangle $,$\left| {{\varphi }_{20}} \right\rangle $ &$\left| {{\varphi }_{4}} \right\rangle $,$\left| {{\varphi }_{14}} \right\rangle $ \\
			&$i=15$ & & & &\\
			&$\left| {{\varphi }_{4}} \right\rangle $,$\left| {{\varphi }_{13}} \right\rangle $ & & & &\\
			\hline
			\multirow{2}{*}{${{a}_{10i}}={{a}_{i10}}=0$} &$i=11$ &$i=12$ &$i=13$ &$i=14$ &$i=15$ \\
			&$\left| {{\varphi }_{8}} \right\rangle $,$\left| {{\varphi }_{13}} \right\rangle $ &$\left| {{\varphi }_{8}} \right\rangle $,$\left| {{\varphi }_{30}} \right\rangle $ &$\left| {{\varphi }_{8}} \right\rangle $,$\left| {{\varphi }_{20}} \right\rangle $ &$\left| {{\varphi }_{8}} \right\rangle $,$\left| {{\varphi }_{14}} \right\rangle $ &$\left| {{\varphi }_{8}} \right\rangle $,$\left| {{\varphi }_{13}} \right\rangle $ \\
			\hline
			\multirow{2}{*}{${{a}_{11i}}={{a}_{i11}}=0$} &$i=12$ &$i=13$ &$i=14$ &$i=15$ & \\
			&$\left| {{\varphi }_{19}} \right\rangle $,$\left| {{\varphi }_{30}} \right\rangle $ &$\left| {{\varphi }_{19}} \right\rangle $,$\left| {{\varphi }_{20}} \right\rangle $ &$\left| {{\varphi }_{19}} \right\rangle $,$\left| {{\varphi }_{14}} \right\rangle $ &$\left| {{\varphi }_{19}} \right\rangle $,$\left| {{\varphi }_{15}} \right\rangle $ &\\
			\hline
			\multirow{2}{*}{${{a}_{12i}}={{a}_{i12}}=0$} &$i=13$ &$i=14$ &$i=15$ & & \\
			&$\left| {{\varphi }_{30}} \right\rangle $,$\left| {{\varphi }_{20}} \right\rangle $ &$\left| {{\varphi }_{30}} \right\rangle $,$\left| {{\varphi }_{14}} \right\rangle $ &$\left| {{\varphi }_{30}} \right\rangle $,$\left| {{\varphi }_{13}} \right\rangle $ & &\\
			\hline
			\multirow{2}{*}{${{a}_{13i}}={{a}_{i13}}=0$} &$i=14$ &$i=15$ & & & \\
			&$\left| {{\varphi }_{20}} \right\rangle $,$\left| {{\varphi }_{14}} \right\rangle $ &$\left| {{\varphi }_{20}} \right\rangle $,$\left| {{\varphi }_{13}} \right\rangle $ & & &\\
			\hline
			\multirow{2}{*}{${{a}_{14i}}={{a}_{i14}}=0$} &$i=15$ & & & & \\
			&$\left| {{\varphi }_{14}} \right\rangle $,$\left| {{\varphi }_{13}} \right\rangle $ & & & & \\
			\hline
			\multirow{6}{*}{${{a}_{00}}={{a}_{ii}}$} & $i=1$ & $i=2$ & $i=3$ & $i=4$ & $i=5$ \\
			&$\left| {{\varphi }_{0,33}} \right\rangle $ &$\left| {{\varphi }_{6,33}} \right\rangle $ &$\left| {{\varphi }_{28,33}} \right\rangle $ &$\left| {{\varphi }_{11,33}} \right\rangle $ &$\left| {{\varphi }_{1,33}} \right\rangle $ \\
			& $i=6$ & $i=7$& $i=8$ &$i=9$ &$i=10$\\
			&$\left| {{\varphi }_{18,33}} \right\rangle $ &$\left| {{\varphi }_{11,33}} \right\rangle $ &$\left| {{\varphi }_{9,33}} \right\rangle $ &$\left| {{\varphi }_{4,33}} \right\rangle $ &$\left| {{\varphi }_{3,33}} \right\rangle $ \\
			&$i=11$ &$i=12$ &$i=13$ &$i=14$ &$i=15$ \\
			&$\left| {{\varphi }_{12,33}} \right\rangle $ &$\left| {{\varphi }_{22,33}} \right\rangle $ &$\left| {{\varphi }_{16,33}} \right\rangle $ &$\left| {{\varphi }_{15,33}} \right\rangle $ &$\left| {{\varphi }_{13,33}} \right\rangle $\\
		\end{tabular}
	\end{ruledtabular}
\end{table*}

Obviously, \textit{BC}’s measurement matrix is proportional to the identity matrix, so it means that \textit{BC}’system starts with a trivial measurement, and cannot get any information about shared state distinction from the measurement result. As for the other two division methods, $AB|C$, $AC|B$, the proof method is similar to this. In summary, the multi-party divided into any two parts can keep the strong nonlocality of the quantum system.$\hfill\blacksquare$

\nocite{*}

\bibliography{UPB.bib}

\end{document}